\documentclass[%
 aip,
 amsmath,amssymb,
 reprint,%
floatfix
]{revtex4-1}

\usepackage{graphicx}
\usepackage{dcolumn}
\usepackage{bm}

\usepackage[utf8]{inputenc}
\usepackage[T1]{fontenc}
\usepackage{mathptmx}
\usepackage{etoolbox}

\makeatletter
\def\@email#1#2{%
 \endgroup
 \patchcmd{\titleblock@produce}
  {\frontmatter@RRAPformat}
  {\frontmatter@RRAPformat{\produce@RRAP{*#1\href{mailto:#2}{#2}}}\frontmatter@RRAPformat}
  {}{}
}%
\makeatother
\begin{document}

\preprint{AIP/123-QED}

\title{Flow of supercooled liquids under dipolar force field}
\author{Kento Maeda}
\affiliation{Graduate School of Arts and Sciences, Meguro-ku, University of Tokyo, Tokyo 153-8902, Japan}

\author{Atsushi Ikeda}
\affiliation{Graduate School of Arts and Sciences, Meguro-ku, University of Tokyo, Tokyo 153-8902, Japan}

\date{\today}

\begin{abstract}
The viscosity of supercooled liquids notably increases with decreasing temperature, leading to solidification through a glass transition. 
This process is accompanied by dynamic heterogeneity, characterized by persistent dynamic spatial correlations. 
This study investigates how dynamic heterogeneity influences the applicability of the Navier-Stokes equations to the flow of supercooled liquids. 
Utilizing molecular dynamics simulations, we subjected a two-dimensional supercooled liquid to a localized dipolar force field and compared the resulting steady velocity field with the prediction from the Navier-Stokes equations. 
Our approach captures a significant breakdown of the Navier-Stokes equations in real space; specifically, supercooled liquids flow more rapidly near the external force than the prediction from the Navier-Stokes equations.
Furthermore, this deviation is enhanced by the supercooling and is accompanied by the growth of dynamic heterogeneity. 
\end{abstract}

\maketitle
\section{Introduction}
When a liquid is rapidly cooled below its freezing point, it often bypasses the usual crystallization process. 
This phenomenon, known as supercooling, is characterized by a significant increase in viscosity as the temperature decreases. 
With further cooling, the supercooled liquid eventually loses its fluidity and solidifies, although the particles remain in a disordered configuration. 
This solidification is referred to as the glass transition, which has been observed in a wide range of systems including silicate glasses, metallic glasses, molecular liquids, polymers, and colloids \cite{Angell1988, Angell1995, Ediger1996review, Debenedetti2001review, Cavagna2009review}.

At the molecular level, the increased viscosity in supercooled liquids is accompanied by a slowdown of structural relaxation. 
As supercooling progresses, the time required for particle rearrangement increases. 
This is evident, for example, in the behavior of the mean-squared displacement and intermediate scattering function. 
In supercooled liquids, these functions typically exhibit the two-step relaxation process. 
The slower phase of this relaxation process, termed $\alpha$-relaxation, strongly depends on temperature. 
As the temperature decreases and approaches the glass transition point, both $\alpha$-relaxation time and viscosity exhibit a divergent increase~\cite{Gotze1992, Yamamoto-Onuki1998, Tong2018, Berthier2011review}. 

In ordinary liquids, particle rearrangement typically occurs at the level of individual particles. 
By contrast, in supercooled liquids, particles tend to cluster together with their neighbors and rearrange themselves collectively as a cluster unit. 
This results in prolonged dynamic spatial correlations in the fluctuations of molecular motion, which is known as dynamic heterogeneity. 
The increase in the correlation length associated with dynamic heterogeneity is believed to be a primary factor in the anomalous behaviors of supercooled liquids and their glass transition \cite{Hurley-Harrowell1995, Kob1997, Yamamoto-Onuki1997, Donati1998, Ediger2000review, Weeks2000, Lacevic2002, Lacevic2003, Widmer2004, Berthier2005, Dalle-Ferrier2007, Flenner-Szamel2010, Flenner2011, Shiba2012, Flenner2014}. 

Macroscopic dynamics of a normal fluid can be described by the Navier-Stokes equations~\cite{Hansen2013}. 
The Navier-Stokes equations are known to be powerful even on a micro-meter scale~\cite{microhydrodynamics}. 
However, the presence of dynamic heterogeneity in supercooled liquids, which exhibit mesoscopic correlation lengths, raises questions about the applicability of the Navier-Stokes equations to supercooled liquids. 
For example, Furukawa and Tanaka investigated the wavenumber dependence of viscosity in supercooled liquids using currents measured in Molecular Dynamics (MD) simulations~\cite{Furukawa2011}. 
They found that the viscosity on microscopic scales widely deviates from the macroscopic one. 
Namely, while the viscosity in the small wavenumber region asymptotically approaches the macroscopic viscosity, the viscosity significantly diminishes in the large wavenumber region. 
This decrease in viscosity in the large wavenumber region becomes more pronounced with advancing supercooling. 
These results imply conventional hydrodynamics is not applicable to supercooled liquids at least on the microscopic scale. 

To our knowledge, however, there is currently no real-space study on the validity of the Navier-Stokes equations in supercooled liquids. 
Our study aims to demonstrate the breakdown of the Navier-Stokes equations in real space. 
To achieve this goal, we draw inspiration from the work of Learner et al.~\cite{Lerner2014} 
They performed MD simulations of amorphous solids undergoing a jamming transition and investigated the applicability of the conventional elasticity theory for amorphous solids~\cite{Lerner2014}. 
Specifically, they applied a localized dipolar force field to a model amorphous solid and compared the resultant displacement field with the prediction from the elasticity theory. 
They observed a significant breakdown of the elasticity theory, which is caused by a spatially disordered displacement field near the external force. 
In this work, we apply a similar approach to supercooled liquids. 
We performed MD simulations on a two-dimensional supercooled liquid subjected to a localized dipolar force field and analyzed the velocity field of the resultant steady flow, aiming to reveal the hydrodynamic property of supercooled liquids~\cite{exfield-hydro2020}. 

The remainder of this paper is organized as follows: Section 2 details the model employed in our simulations. 
Section 3 characterizes the dynamics of the system at equilibrium, confirming that our model is a fragile supercooled liquid. 
Section 4 introduces a method for analyzing the velocity field induced by an external dipolar force field and demonstrates how the supercooled liquid deviates from the Navier-Stokes equations. 
This section also discusses a connection between the breakdown of the Navier-Stokes equations and dynamic heterogeneity. 
Section 5 summarizes the results and discusses future research directions.

\section{Model}
We consider $N$ particles in a two-dimensional simulation box. 
The coordinate and velocity of particle $i$ are denoted by $\vec{r}_i = (x_i, \: y_i)$ and $\vec{u}_i=(u_{xi}, \: u_{yi})$, respectively. 
The system is a binary mixture of 8,000 large particles and 8,000 small particles, hence $N=16000$. 
The particles interact via the Weeks-Chandler-Andersen (WCA) potential~\cite{WCA2003}:
\begin{equation}
v(r_{ij})=
\left\{
\begin{array}{ll}
4 \epsilon \left[ \left(\frac{\sigma_{ij}}{r_{ij}}\right) ^{12}-\left(\frac{\sigma_{ij}}{r_{ij}}\right)^6  \right]+ \epsilon & \left(\frac{r_{ij}}{\sigma_{ij}} \le 2^{\frac{1}{6}}\right) \\
0 & (\mathrm{otherwise})
\end{array}
\right.
,
\end{equation}
where $r_{ij}$ denotes the distance between particles $i$ and $j$, and $\sigma_{ij}$ is the mean diameter of particles $i$ and $j$. 
The ratio of the diameters of large and small particles is set to be $\sigma_l/\sigma_s = 1.4$. 
We set the mass of each particle to $m = 1$ irrespective of its size and set $\sigma_s = 1$ and $\epsilon = 1$, which determine the units in this work. 
The simulation box is square with a side length of $L=126.49$, which fixes the number density of the system at 1. 
The volume of the system is $V=L^2$. 
Periodic boundary conditions are imposed on the system.

For this model, we performed MD simulations. 
The velocity-Verlet method was applied to integrate Newton's equations of motion with a time step $0.002$. 
Temperature $T$ is measured in $\epsilon /k_\mathrm{B}$, where $k_\mathrm{B}$ is the Boltzmann's constant.

\section{Equilibrium dynamics}
In this section, we present the results of MD simulations of our model in equilibrium states. 
The results establish that our model exhibits a typical behavior of fragile supercooled liquids.  
The viscosities computed in this section will be used to analyze the flow of supercooled liquids in the next section. 

\subsection{Method}
For the equilibrium simulations, we first performed equilibration runs at the target temperatures $T=19$, 15, 12, 11, and 10 in the NVT ensemble. 
The temperature was controlled using the canonical sampling velocity rescaling (CSVR) method~\cite{CSVR2007}. 
Note that the time duration for the equilibration runs ($t=10^3$) is sufficiently longer than the relaxation time of the system. 
Then starting from the equilibrium configurations, we performed production runs in the NVE ensemble. 
For a single target temperature, we performed several production runs starting from independent initial configurations to improve the statistics.
Hereafter, $\langle \cdots \rangle$ denotes the average over the independent samples and/or the initial time.  

\subsection{Results}
We first calculated the radial distribution function $g(r)$ at the target temperatures and confirmed that $g(r)$ is insensitive to temperature and does not show any sign of crystallization in the temperatures studied. 
We then calculated the mean-squared displacement, overlap function, and stress autocorrelation function to characterize the equilibrium dynamics of the system.

\subsubsection{Mean-squared displacement}
\label{section_MSD}

\begin{figure}[t]
\includegraphics[width=\columnwidth]{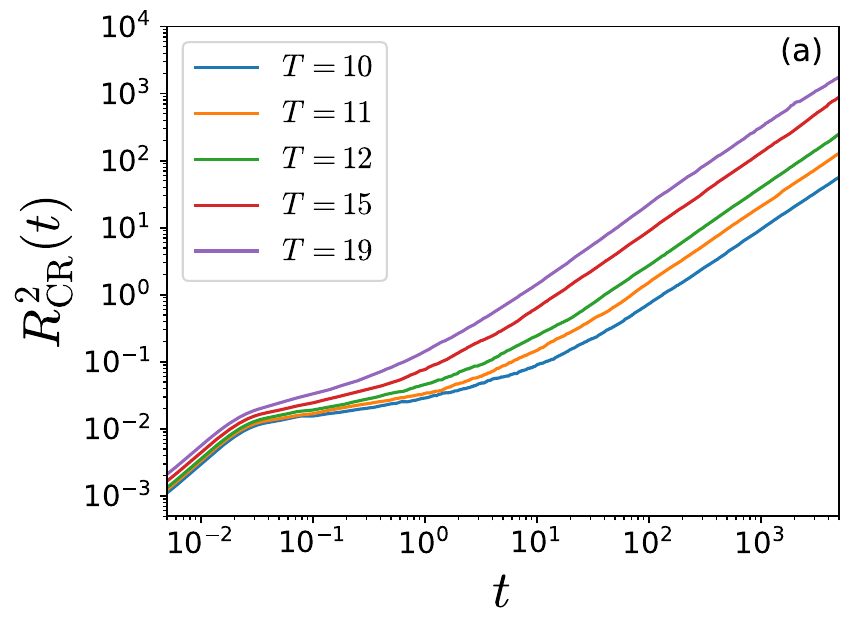}
\includegraphics[width=\columnwidth]{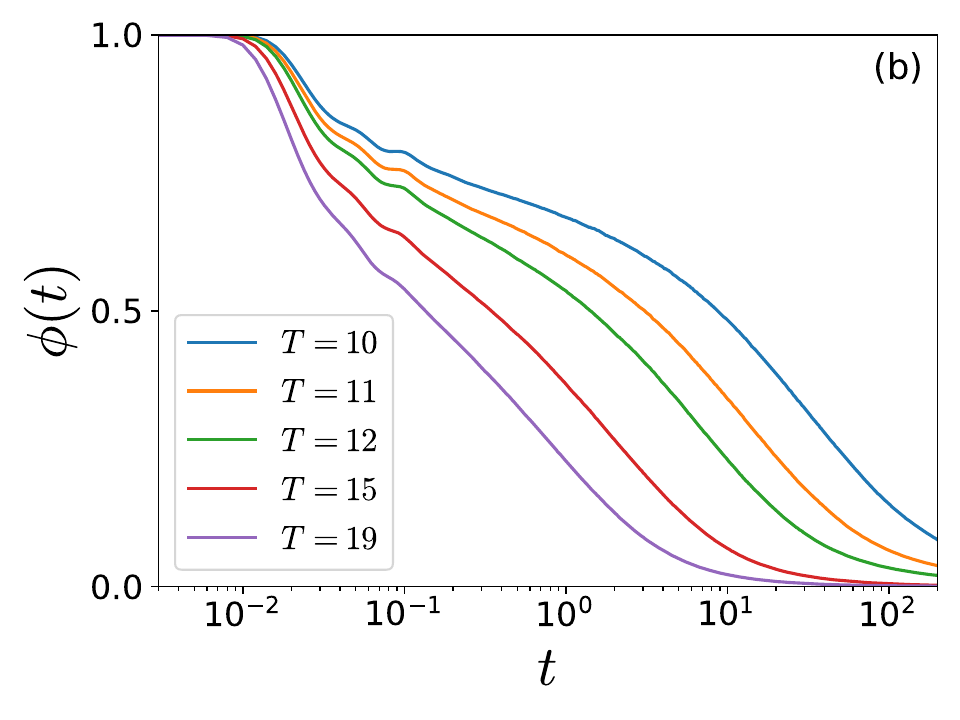}
\caption{(a) The cage-relative mean-squared displacement (MSD) at the target temperatures in the equilibrium states. 
(b) The cage-relative overlap function at the target temperatures in the equilibrium states. }
\label{fig_msd_and_bond_breakage}
\end{figure}

Let $\Delta \vec{r}_i(t)$ denote the displacement of particle $i$ from $t=0$ to $t=t$: $\Delta \vec{r}_i(t) = \vec{r}_i(t)-\vec{r}_i(0)$. 
The cage-relative displacement of particle $i$, represented by $\Delta \vec{r}_{\mathrm{CR}, i}(t)$, is defined as:
\begin{equation}
\Delta \vec{r}_{\mathrm{CR}, i}(t)=\Delta \vec{r}_i(t) - \frac{1}{N_i^\mathrm{n.n.}}\sum_{j \in \mathrm{n.n.}} \Delta \vec{r}_j(t) .
\label{def_CR_displacement}
\end{equation}
``$\mathrm{n.n.}$'' in the summation represents the nearest neighbors of the particle $i$ and $N_i^\mathrm{n.n.}$ is the number of them. 
In this study, particle $j$ is considered a nearest neighbor of particle $i$ when $|\vec{r}_i(0) - \vec{r}_j(0)| < 1.25 \sigma_{ij}$. 
Note that 1.25 roughly corresponds to the first minimum of the radial distribution function $g(r)$. 
The cage-relative displacements are useful to observe the inherent relaxation dynamics without being influenced by the long-wavelength Mermin-Wagner fluctuations in two-dimensional systems~\cite{Shiba2016, Bernd2017, Flenner-Szamel2019, Li2019}. 

The cage-relative mean square displacement (MSD), defined by
\begin{equation}
R_{\mathrm{CR}}^2(t) = \left \langle \frac{1}{N} \sum_{i=1}^N |\Delta \vec{r}_{\mathrm{CR}, i}(t)|^2  \right \rangle ,
\end{equation}
estimates the average distance a particle has moved from its initial coordinates. 
In Fig.~\ref{fig_msd_and_bond_breakage}(a), we present the MSD $R_{\mathrm{CR}}^2(t) $ at the target temperatures. 
In all cases, the system exhibits ballistic behavior, characterized by $R_{\mathrm{CR}}^2(t)  \propto t^2$, on a short timescale of $t \lesssim 10^{-2}$. 
The ballistic behavior transitions to the diffusive behavior, with $R_{\mathrm{CR}}^2(t) \propto t$, on a longer timescale.
At lower temperatures, this transition takes a longer time, resulting in the emergence of the plateau of the MSDs. 
This is a characteristic feature commonly observed in supercooled liquids, which is attributed to the cage effect: particles require a long time to escape from the cage formed by their surrounding particles~\cite{Cage-Effect1998}. 
Note that although long-wavelength Mermin-Wagner fluctuations occur in two-dimensional systems, the cage-relative displacement helps reduce the effect of these fluctuations and allows observation of plateaus in the MSDs.~\cite{Shiba2016, Bernd2017, Flenner-Szamel2019, Li2019}.

\subsubsection{Overlap function}
\label{section_overlap}

We next computed the cage-relative overlap function $\phi(t)$. 
The cage-relative overlap function is defined as~\cite{Shiba2018}: 
\begin{equation}
D_i(t)=\Theta(a-|\Delta \vec{r}_{\mathrm{CR}, i}(t)|) ,
\end{equation}
where $\Theta(x)$ is Heaviside's step function and we set $a=0.15$. 
We then define a time-correlation function $\phi(t)$ as
\begin{equation}
\phi(t)= \left \langle \frac{1}{N}\sum_{i=1}^N D_i(t)\right \rangle,
\end{equation}
which quantifies the proportion of particles whose cage-relative displacement remains below the threshold $a$. 

We plotted the overlap function $\phi(t)$ against the logarithm of time in Fig.~\ref{fig_msd_and_bond_breakage}(b). 
As the temperature is lowered, the structural relaxation becomes increasingly slower and the overlap function displays a two-step relaxation, which is a characteristic feature of supercooled liquids. 
The relaxation time $\tau$, defined by $\phi(t = \tau) = e^{-1}$, increases noticeably as the temperature decreases.

\subsubsection{Stress autocorrelation function}
\label{section_visc}
\begin{figure}[t]
\includegraphics[width=\columnwidth]{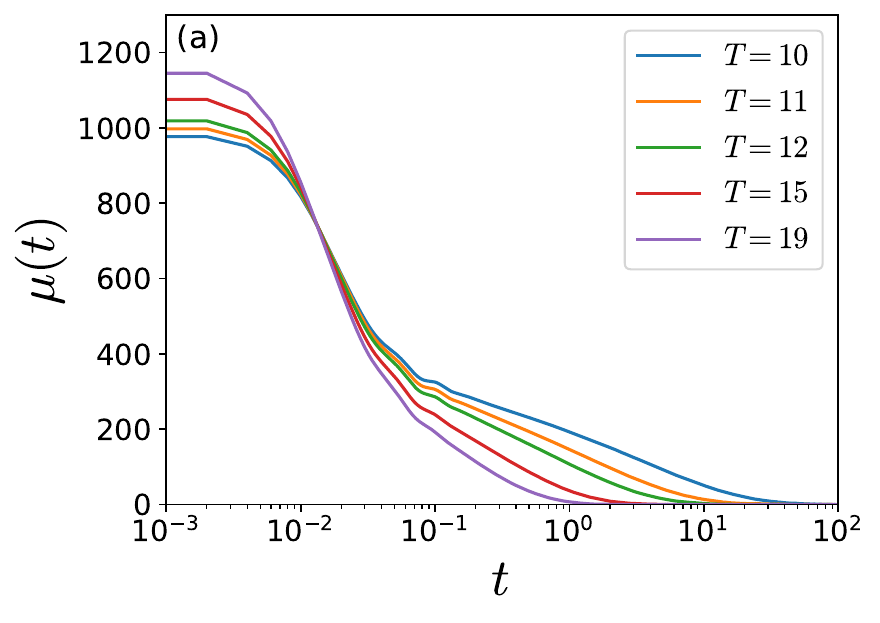}
\includegraphics[width=\columnwidth]{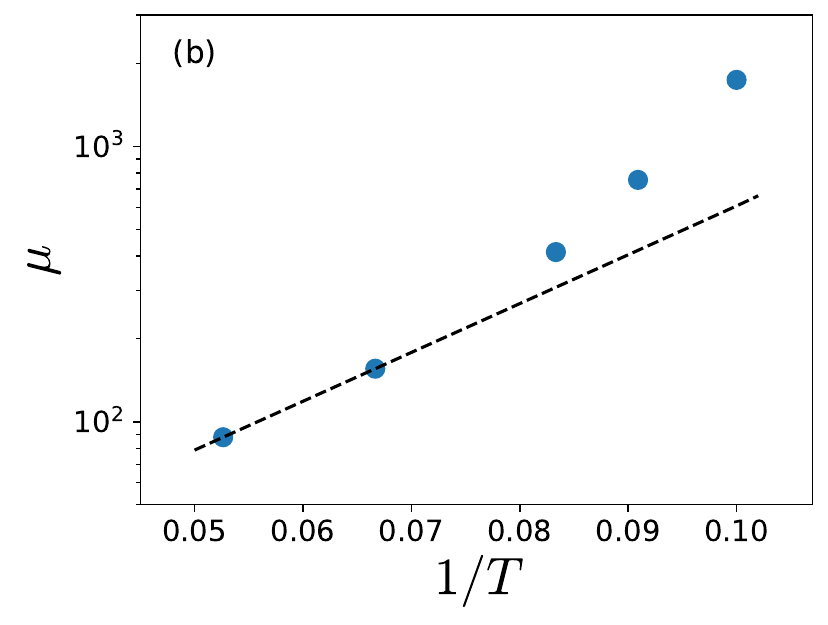}
\caption{(a) The time correlation function of the shear stress at the target temperatures in the equilibrium states. 
(b) The shear viscosity is plotted against inverse temperature. 
The dashed line indicates the Arrhenius behavior at high temperatures.}
\label{fig_viscosity}
\end{figure}

We finally calculated the time-correlation function of the shear stress
\begin{equation}
\mu(t) = \frac{1}{VT} \langle \Pi^{xy}(t) \Pi^{xy} \rangle
\end{equation}
at the target temperatures. 
The shear stress $\Pi^{xy}$ is given by 
\begin{equation}
\Pi^{xy} = \sum_{i=1}^N \left( m u_{ix} u_{iy} + \frac{1}{2} \sum_{j \neq i}^N \frac{x_{ij} y_{ij}}{r_{ij}} v'(r_{ij})  \right), 
\end{equation}
with $x_{ij}=x_j-x_i$ and $y_{ij}=y_j-y_i$~\cite{Hansen2013}. 

The results are shown in Fig.\ref{fig_viscosity}(a). 
As in the case of the overlap function $\phi(t)$ shown in Fig.~\ref{fig_msd_and_bond_breakage}(b), the relaxation becomes increasingly slower as the temperature is lowered. 
The slow relaxation of the stress autocorrelation function indicates that the system is more likely to exhibit solid-like mechanical properties at lower temperatures. 
$\mu(t)$ exhibits a two-step relaxation similar to that observed for $\phi(t)$, which is again a characteristic feature of supercooled liquids~\cite{Varnik2006}.

The macroscopic shear viscosity $\mu$ of the system was computed using the Green-Kubo relations:
\begin{equation}
\mu = \int_0 ^ \infty \mu(t) dt .
\end{equation}
In practice, we first calculated accurate $\mu(t)$ by averaging over 10 samples at each temperature and then utilized the trapezoidal rule for the numerical integration. 
To suppress the numerical error, we terminated the numerical integration when the correlation became negative for the first time. 
In Fig.~\ref{fig_viscosity}(b), we present the obtained shear viscosity $\mu$ against the inverse temperature $1/T$. 
Clearly, the viscosity increases in a super-Arrhenius manner at low temperatures. 
We therefore conclude that our system exhibits the characteristic dynamics of fragile glass-forming liquids. 
The viscosities obtained in this section are used in the next section. 

\section{Dynamics under the dipolar force field}
\begin{figure}
  \centering
  \includegraphics[width=\columnwidth]{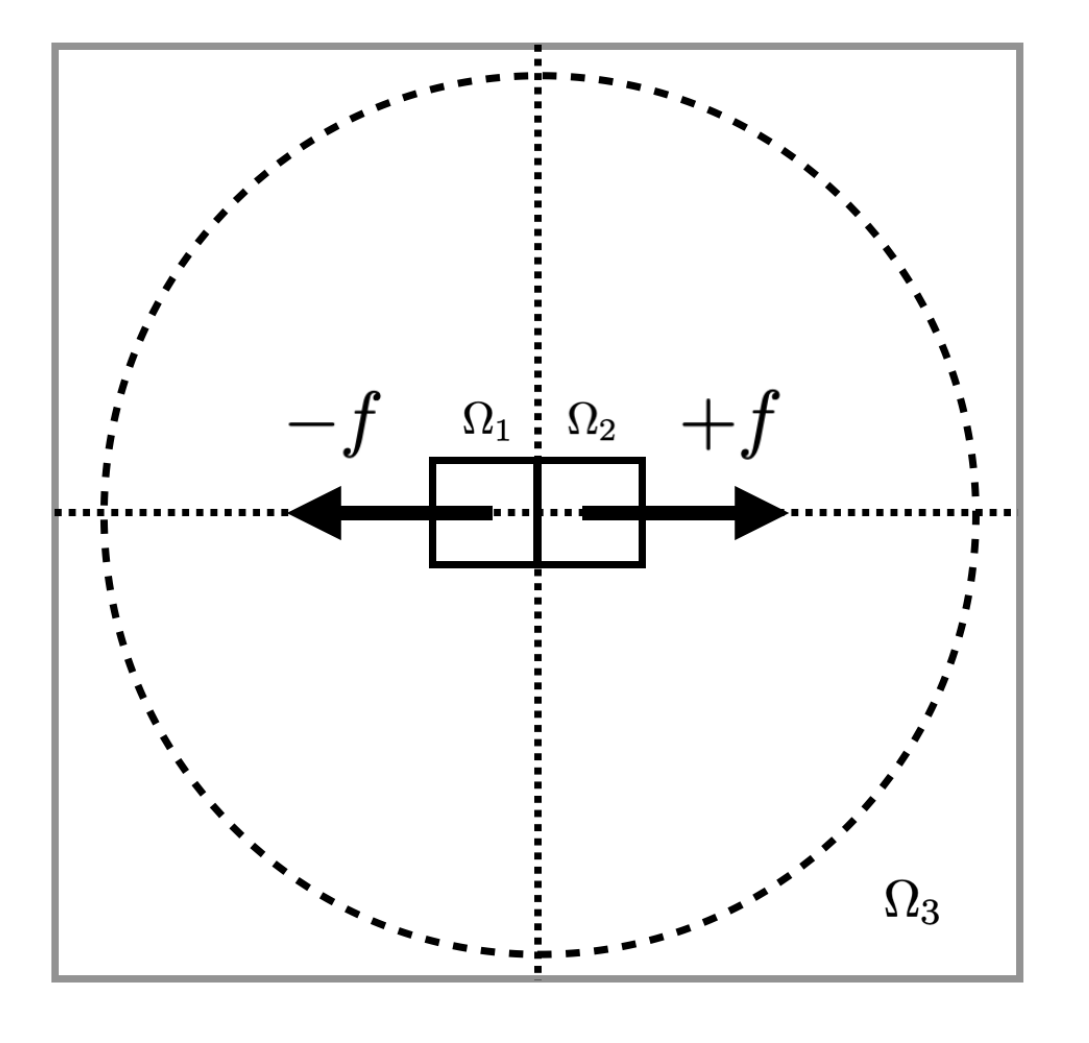}
  \caption{Scheme of simulations of the dynamics under a dipolar force field.}
  \label{scheme}
\end{figure}

In this section, we perform MD simulations of our model under an external dipolar force field. 
We investigate the microscopic and macroscopic dynamics in a steady flow state to evaluate the applicability of the Navier-Stokes equations to supercooled liquids. 
Furthermore, we discuss the structural relaxation under the dipolar force field.

\subsection{Method}
We applied a small external force field to our two-dimensional system and probed the resultant steady flow. 
To ensure force balance throughout the entire system, we introduced a dipolar force field, consisting of two force fields of equal magnitude but opposite direction. 
Specifically, we defined two small regions at the center of the system and applied the force fields in these regions as follows. 
We introduce the $xy$ coordinate system with its origin at the center of the system, and define the two regions $\Omega_1$ and $\Omega_2$ as shown in Fig. \ref{scheme}:
\begin{equation}
\begin{split}
\Omega_1 &= \{ (x,\:y) \: | \: -2<x<0, \:-1<y<1 \}, \\ 
\Omega_2 &= \{ (x,\:y) \: | \: 0<x<2, \:-1<y<1 \}.
\end{split}
\end{equation}
The numbers of particles in $\Omega_1$ and $\Omega_2$, denoted as $N_1$ and $N_2$ respectively, were counted at each time step.
We then apply the following external force to each particle $i$:
\begin{equation}
\vec{F}_i^{\mathrm{ext}} = 
\left\{
\begin{array}{ll}
-\frac{f}{N_1} \vec{e}_x &(\vec{r}_i \in \Omega_1) \\
\frac{f}{N_2}  \vec{e}_x &(\vec{r}_i \in \Omega_2) \\
0 & (\mathrm{otherwise})
\end{array}
\right. , 
\end{equation}
where $\vec{e}_{\alpha}$ is the unit vector along the direction $\alpha$. 
We note that although the values of $N_1$ and $N_2$ may vary over time, the external force $\vec{F}_i^{\mathrm{ext}}$ ensures the conservation of the system's total momentum. 
The magnitude of the external force $f$ is constant throughout the simulations, and we examined five different cases of $f=50, 80, 100, 200$, and $500$. 

The work done by the external forces could heat the overall system. 
To prevent this, the periphery of the simulation box, denoted as $\Omega_3$, was coupled to a heat bath maintained at temperature $T$ with the CSVR method. 
Specifically, $\Omega_3$ was defined as the region outside a circle, centered within the simulation box, with a radius of $L/2 - 2$ (illustrated in Fig. \ref{scheme}). 
We define the number of particles in $\Omega_3$ as $N_3$ and the average velocity of particles in $\Omega_3$ as 
\begin{equation}
\vec{V}_3 = \frac{1}{N_3} \sum_{i : \: \vec{r}_i \in \Omega_3} \vec{u}_i ,
\end{equation}
where $N_3$ may also vary over time. 
For particles in $\Omega_3$, we calculate the velocity deviation from the average
\begin{equation}
\vec{v}_i = \vec{u}_i - \vec{V}_3,  
\end{equation}
and the associated temperature 
\begin{equation}
T_{3\mathrm{before}} = \frac{1}{N_3} \sum_{i : \: \vec{r}_i \in \Omega_3} |\vec{v}_i|^2. 
\end{equation}
Then, following the CSVR method, we updated the velocity of particles in $\Omega_3$ by 
\begin{equation}
\vec{u}_i \to \vec{V}_3+\sqrt{\frac{T_{3\mathrm{after}}}{T_{3\mathrm{before}}}} \vec{v}_i, 
\end{equation}
where the updated temperature is given by 
\begin{equation}
T_{3\mathrm{after}} = \frac{1}{N_3} \sum_{i=1}^{N_3} |\vec{v}_i^{\mathrm{M}}|^2 ,
\end{equation}
where $\vec{v}_i^{\mathrm{M}}$ is a random number sampled from a Maxwell distribution at temperature $T$.

Using the method outlined above, we simulate the steady flow of supercooled liquids induced by a dipolar external force field while avoiding the temperature increase. 
We first performed preparatory runs with the external force. 
Starting from the equilibrium configurations obtained in Sec.~III, we perform the MD simulations outlined above for a duration of $t=1000$. 
The preparatory runs ensure that the system reaches a steady flow state.
Then, starting from the configurations obtained in the preparatory runs, we perform the same MD simulations as in the production runs. 
The time duration of the production runs is $t=10^5$. 
We analyzed the velocity field of the model during these production runs. 

\subsection{Analysis of Steady Flow}

To clarify the gap between the macroscopic and microscopic perspectives of steady flow, this subsection presents the theoretical prediction from hydrodynamics and the description based on MD simulations. 

\subsubsection{Hydrodynamics}

In hydrodynamics, the velocity field $\vec{u}_{\mathrm{NS}}(\vec{r})$ is described by the Navier-Stokes equations. 
When the external force $\vec{f} (\vec{r})$ is sufficiently weak, the fluid is in the Stokes regime, and the flow is described using the Oseen tensor $\mathbb{G}(\vec{r})$:
\begin{equation}
\vec{u}_{\mathrm{NS}}(\vec{r}) = \int d\vec{\xi}\: \mathbb{G}(\vec{r}-\vec{\xi}) \cdot \vec{f}(\vec{\xi}) .
\label{eq_umacro1}
\end{equation}
In the two-dimensional fluid, the Oseen tensor $\mathbb{G}(\vec{r})$ is given by:
\begin{equation}
G_{\alpha \beta}(\vec{r}) = \frac{1}{4\pi \mu r}\left(\delta_{\alpha \beta} \left( -\log r -\frac{1}{2}\right)+\frac{r_\alpha r_\beta}{r^2} \right),
\label{eq_Oseen}
\end{equation}
where $\alpha$ and $\beta$ are the Cartesian components $x$ or $y$, respectively, $r = \sqrt{x^2 + y^2}$, and $\mu$ is the viscosity of the liquid.
Expanding $\mathbb{G}(\vec{r}-\vec{\xi})$ around $\vec{\xi} = \vec{0}$ yields:
\begin{equation}
G_{\alpha \beta}(\vec{r}-\vec{\xi}) = \sum_{n=0}^\infty \frac{(-1)^n}{n!} (\vec{\xi}\cdot \vec{\nabla})^n G_{\alpha \beta}(\vec{r}) .
\label{eq_expansion_Oseen}
\end{equation}

Neglecting the fluctuations of the number density, the force field employed in our simulations can be written as:
\begin{equation}
\vec{f} (\vec{r}) = -\frac{f}{4}\vec{e}_x \mathbf{1}_{\Omega_1}(\vec{r})+\frac{f}{4}\vec{e}_x \mathbf{1}_{\Omega_2}(\vec{r}) .
\label{ext_field}
\end{equation}
Here, $\mathbf{1}_{\Omega_1}(\vec{r})$ and $\mathbf{1}_{\Omega_2}(\vec{r})$ are indicator functions for the set $\Omega_1$ and $\Omega_2$, respectively, defined by:
\begin{equation}
\mathbf{1}_{\Omega_i}(\vec{r}) =
\left\{
\begin{array}{ll}
1 & \left( \vec{r} \in \Omega_i \right) \\
0 & (\mathrm{otherwise})
\end{array}
\right.
,
\end{equation}
with $i=1, 2$. The denominator 4 in Eq. \eqref{ext_field} serves as a normalization factor corresponding to the areas of $\Omega_i$. 
Upon substituting the expansion of the Oseen tensor and the force field expression into the equation for $\vec{u}_\mathrm{NS}(\vec{r})$, we observe that the $n=0$ term from Eq. \eqref{eq_expansion_Oseen} cancels out, leaving the $n=1$ term as the primary contribution. The $n=1$ term's contribution to $\vec{u}_\mathrm{NS}(\vec{r})$ is
\begin{equation}
\vec{u}_{\mathrm{NS}}^{(1)}(\vec{r})=\frac{f}{2\pi \mu r} \cos(2\theta) \hat{r} .
\label{eq_umacro2}
\end{equation}
Here, we introduced polar coordinates $\vec{r}=(r, \theta)$ with its origin at the center of the system, and $\hat{r}=\vec{r}/|\vec{r}|$ denotes the unit vector in the direction of $\vec{r}$. 

Hereafter, we focus on the following quantity 
\begin{equation}
I(r)=\frac{1}{2\pi}\int d\theta \vec{u}(\vec{r}) \cdot \hat{r} \cos(2\theta) .
\label{eq_def_I}
\end{equation}
$I(r)$ quantifies the amplitude of the $n=1$ component of the flow at a radial distance $r$. 
We call $I(r)$ flow descriptor.  
Substituting Eq. \eqref{eq_umacro2} into Eq. \eqref{eq_def_I}, we obtain the prediction by the Navier-Stokes equations for $I(r)$: 
\begin{equation}
I_{\mathrm{NS}}(r)=\frac{1}{2\pi}\int d\theta \vec{u}_{\mathrm{NS}}^{(1)}(\vec{r}) \cdot \hat{r} \cos(2\theta) = \frac{f}{4\pi \mu r}  .
\label{eq_def_INS}
\end{equation}
This result indicates that the steady flow induced by a dipolar force field diminishes inversely with the radial distance in the hydrodynamic theory. 
The higher order terms beyond $n=1$ are deemed irrelevant, which will be demonstrated numerically in Sec. \ref{breakdown}.

\subsubsection{MD Simulations}

Within the microscopic framework, the velocity field $\vec{u}_\mathrm{MD}(\vec{r})$ is directly extracted from MD simulations:
\begin{equation}
\vec{u}_\mathrm{MD}(\vec{r}) = \left\langle \sum_i^N \vec{u}_i(t) \delta(\vec{r}-\vec{r}_i(t)) \right\rangle 
\label{def_microu}
\end{equation}
By substituting the microscopic velocity field Eq. \eqref{def_microu} into Eq. \eqref{eq_def_I}, we obtain the simulation-based flow descriptor $I_\mathrm{MD}(r)$, expressed as:
\begin{equation}
\begin{aligned}
I_\mathrm{MD}(r) &= \frac{1}{2\pi}\int d\theta \vec{u}_{\mathrm{MD}} (\vec{r}) \cdot \hat{r} \cos(2\theta) \\
&=  \frac{1}{2\pi} \left\langle \int d\theta \sum_i^N \vec{u}_i(t)  \cdot \hat{r} \cos(2\theta) \delta(\vec{r}-\vec{r}_i(t)) \right\rangle \\
&=  \frac{1}{2\pi} \left\langle \frac{1}{r \Delta r} \sideset{}{'} \sum_i  \vec{u}_i(t)  \cdot \hat{r}_i \cos(2\theta_i) \right\rangle .
\end{aligned}
\label{eq_I_MD}
\end{equation}
In the final line, we discretized the radial distance $r$ with the width $\Delta r$ to obtain the expression for the practical calculations in the MD simulations. 
$\sideset{}{'}\sum_i$ represents the sum over particles satisfying $r-\frac{1}{2}\Delta r \le |\vec{r}_i| < r+\frac{1}{2}\Delta r$, and $\hat{r}_i = \vec{r}_i/|\vec{r}_i|$. 
Note that $r$ in the term $\frac{1}{r \Delta r}$ in Eq. \eqref{eq_I_MD} is the Jacobian determinant associated with the integration in the polar coordinate system.

The averaging process represented by $\langle \cdots \rangle$ in Eqs. \eqref{def_microu} and \eqref{eq_I_MD} is important to average out the thermal velocities of particles and isolate the flow velocities of particles. 
We do this by both the time averaging and the sample averaging. 
In the time averaging, we recorded physical quantities during the production runs with the time duration $t = 10^5$ and then averaged them. 
In the sample averaging, we performed a large number of independent production runs starting from different initial configurations and then averaged the obtained results. 
The number of independent production runs for this averaging was determined based on the stability of the results. 
For example, for $T=10$ and $f=100$, we needed 3000 independent production runs to ensure the reliability of the numerical results.

\subsection{Linear response}

We aim to compare $I_\mathrm{NS}(r)$ and $I_\mathrm{MD}(r)$ to examine the validity of the hydrodynamic description of the flow of supercooled liquids. 
To this end, we need to calculate $I_\mathrm{MD}(r)$ in the linear response regime of weak external force $f$ since we focused on the Stokes regime in the hydrodynamics to obtain $I_\mathrm{NS}(r)$. 
This subsection is devoted to identifying the linear response regime in the MD simulations. 

\begin{figure}
  \centering
  \includegraphics[width=\columnwidth]{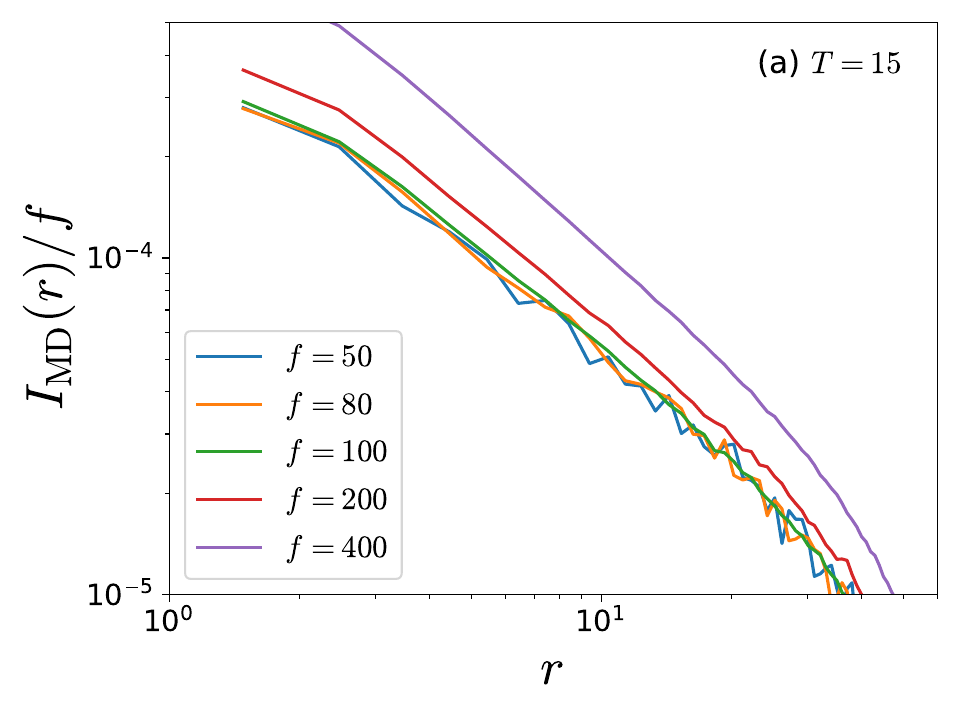}
  \includegraphics[width=\columnwidth]{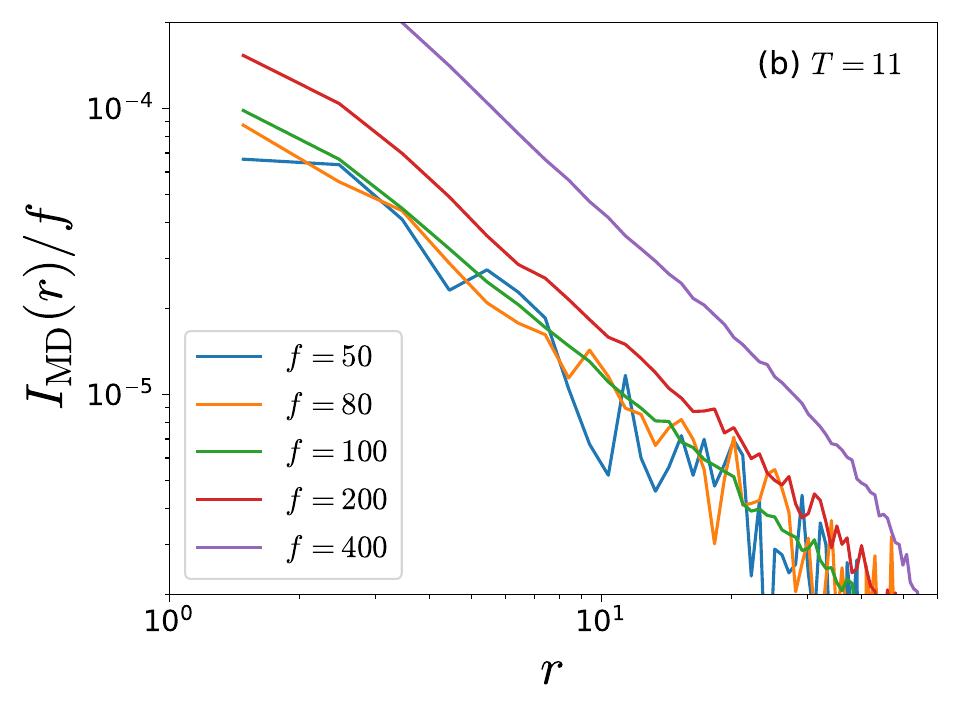}
  \caption{Measurements of $I_{\mathrm{MD}}(r)/f$ for two cases, (a) temperature $T=15$ and (b) $T=11$, while varying the magnitude of the external force $f$.}
  \label{fig_linres}
\end{figure}

To achieve this, we focus on the quantity $I_\mathrm{MD}(r)/f$, which should be independent of $f$ in the linear response regime. 
Fig. \ref{fig_linres} shows $I_\mathrm{MD}(r)/f$ at $T=15$ and at $T=11$. 
The former case is almost in a normal fluid regime while the latter case is in a supercooled regime, at which the MSD exhibits the plateau (see Fig.~\ref{fig_msd_and_bond_breakage}(a)). 

At $T=15$ (Fig. \ref{fig_linres}(a)), the curves for $f=400$ and $200$ are notably higher than the rest, indicating deviations from the linear response. 
In contrast, the curves for $f=100$, $80$, and $50$ are nearly indistinguishable, suggesting that these systems are in the linear response regime. 
This observation is valid also at the lower temperature $T=11$ (Fig. \ref{fig_linres}(b)). 
The curves for $f=100$, $80$, and $50$ are nearly indistinguishable. 
These results suggest that the system at $f \lesssim 100$ is in the linear response regime at various temperatures including the supercooled regime. 
We confirmed that this criteria is valid in all the studied temperatures $T \geq 10$. 

We also observe that the data scatter more for smaller $f$. 
This is because the steady flow becomes slower at smaller $f$ and then we need an averaging over a larger sample size in Eq.\eqref{eq_I_MD} to isolate the flow velocities from the thermal velocities. 
Therefore to minimize the statistical errors, we use $I_{\rm MD}(r)$ obtained at $f = 100$ in the following analysis. 

We note that $f = 100$ is comparable to the typical interparticle force in the system. 
We quantified the typical magnitude of the interparticle force $F_\mathrm{typ}=\langle |v'(r_{ij})| \rangle_{r<\sigma_{ij}}$, where $\langle \cdots \rangle_{r<\sigma_{ij}}$ denotes the average over all particle pairs $i, j$ for which the distance is shorter than $\sigma_{ij}$, and $v'(r)$ represents the derivative of the potential with respect to the distance. 
We found $F_\mathrm{typ}=279$ at $T=19$, meaning that an external force of $f=100$ is slightly smaller than the typical interparticle forces.

\subsection{Breakdown of the Navier-Stokes equations}
\label{breakdown}

We now compare the simulation-based flow descriptor $I_\mathrm{MD}(r)$ with the hydrodynamic prediction $I_{\mathrm{NS}}(r)$ to determine whether the flow of supercooled liquids follows the Navier-Stokes equations. 
For $I_\mathrm{MD}(r)$, we focus on the results with the external force $f = 100$. 
For $I_\mathrm{NS}(r)$, we use the formula Eq.\eqref{eq_def_INS} where the value of viscosity $\mu$ is taken from the MD simulation results illustrated in Fig.~\ref{fig_viscosity}(b). 

\begin{figure}
  \centering
  \includegraphics[width=\columnwidth]{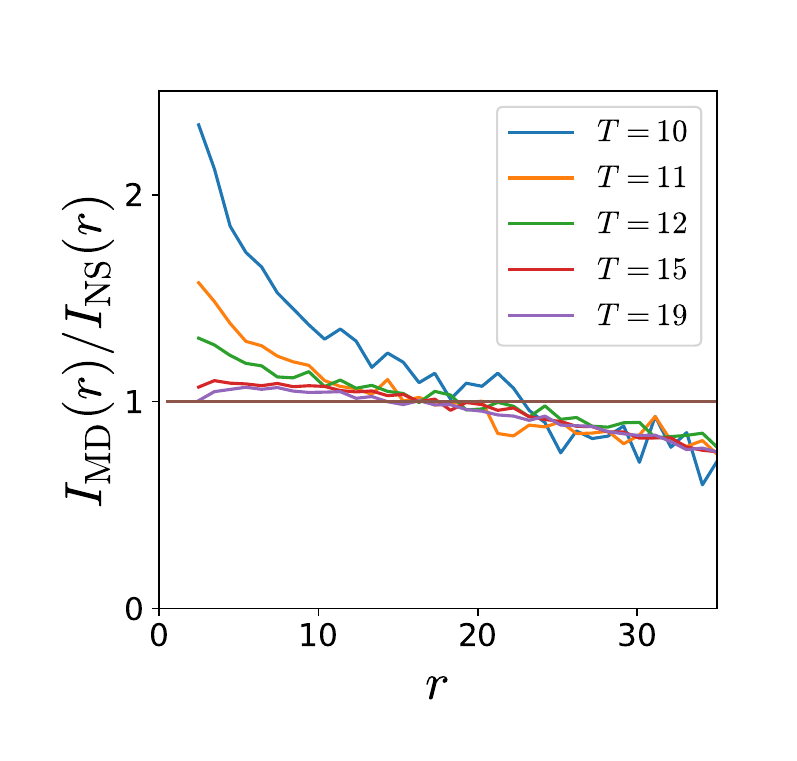}
  \caption{The value of the ratio of ${I_\mathrm{MD}(r)}$ to ${I_\mathrm{NS}(r)}$ at each temperature. A larger value than 1 indicates a flow velocity greater than that predicted by the Navier-Stokes equations.}
  \label{main_fig}
\end{figure}

Fig.~\ref{main_fig} displays $I_\mathrm{MD}(r)$ normalized by $I_\mathrm{NS}(r)$ at the target temperatures. 
At higher temperatures, such as $T=19$, the ratio $\frac{I_\mathrm{MD}(r)}{I_\mathrm{NS}(r)}$ is close to $1$ over a wide range of $r$. 
Interestingly, $\frac{I_\mathrm{MD}(r)}{I_\mathrm{NS}(r)} \simeq 1$ works well even for the very short length $r \simeq 2$. 
Given that the unit length 1 is approximately the particle size, this implies that macroscopic hydrodynamics is applicable down to the particle scale for high-temperature liquids.

We now focus on the low-temperature results. 
With decreasing temperature, $\frac{I_\mathrm{MD}(r)}{I_\mathrm{NS}(r)}$ becomes noticeably larger than 1 at the short distance. 
This indicates that the flow near the external force is enhanced beyond the prediction from the Navier-Stokes equations. 
This deviation becomes more pronounced at lower temperatures; in deeply supercooled states, the flow velocity field increasingly deviates from the hydrodynamic prediction. 
We emphasize that $I_{\rm MD}(r)$ is calculated in the linear response regime as we showed in the previous subsection. 
Therefore, these results establish that supercooled liquids near the external force flow more rapidly than the hydrodynamic prediction. 

Note that the ratio $\frac{I_\mathrm{MD}(r)}{I_\mathrm{NS}(r)}$ falls below 1 at approximately $r=20$ although this ratio is expected to approach 1 as $r \to \infty$.
This discrepancy is due to the boundary effect.
The system is attached to the thermostat in the region $r \geq L/2 -2 = 61$, and $r=20$ is about $30\%$ of this cutoff length. 
To estimate the boundary effect, we performed simulations with the quadrupled particle number $4N$ and the doubled box size $2L$ and found that $\frac{I_\mathrm{MD}(r)}{I_\mathrm{NS}(r)}$ falls below 1 at around the $30\%$ of the new cutoff length. 
This means that the influence of the boundary conditions is negligible in the region of $r \lesssim 20$ for the original system.

Within hydrodynamics, our results imply that viscosity decreases at shorter length scales. 
This translates into a wavenumber-dependent viscosity that decreases at higher wavenumbers.
Previous assessments of viscosity in wavenumber space have shown a decrease from macroscopic values in high-wavenumber regions~\cite{k-dependent-visc-Kim2005, k-dependent-visc-Furukawa2009, k-dependent-visc-Puscasu2010, Furukawa2011}. 
Our results are in qualitative agreement with these observations.

\subsection{Structural relaxation in the flowing supercooled liquids}
\begin{figure}
  \centering
    \includegraphics[width=\columnwidth]{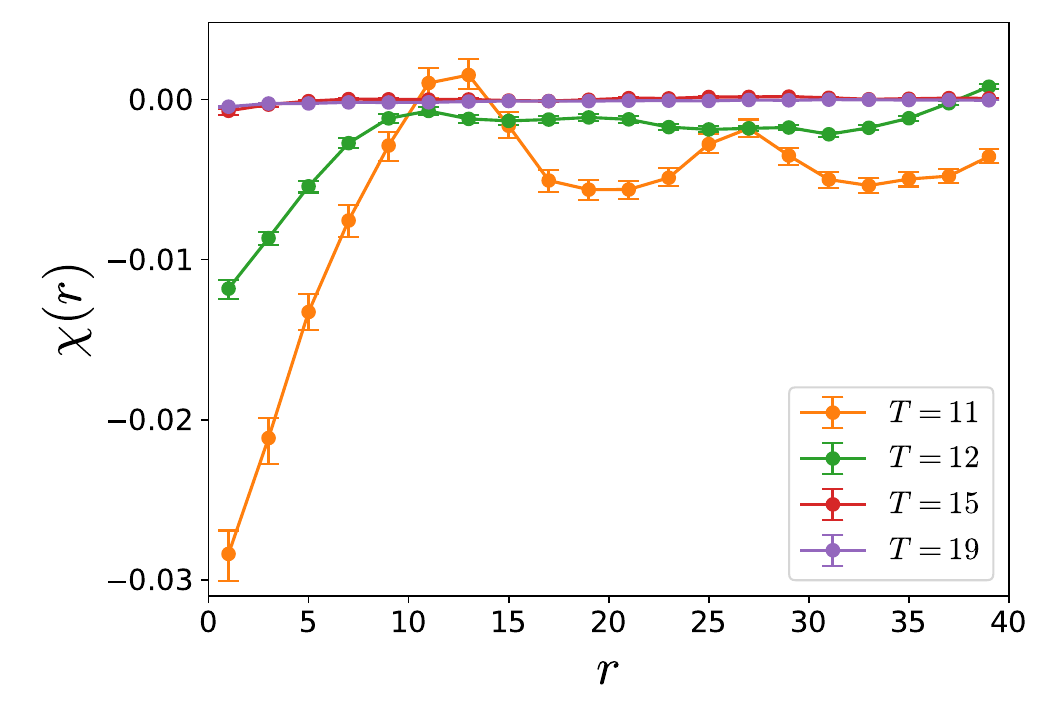}
  \caption{The ``susceptibility'' of the relaxation time  $\chi (r)$. 
The results are measured at $T = 11, 12, 15, 19$ with a constant force $f = 100$.}
  \label{fig_chi}
\end{figure}

In supercooled liquids, microscopic relaxation time is critical in determining viscosity \cite{Varnik2006}. 
The results in the previous section would suggest that the microscopic relaxation times also decrease in proximity to the external force. 
To test this hypothesis, we measured the cage-relative overlap function in the steady flow states. 

In our setting of the steady flow, the overlap function can depend on the distance from the external force $r$ and the amplitude of the external force $f$; hence, the overlap function is denoted by $\phi(t;r,f)$. 
We then define the relaxation time in the steady flow states $\tau (r,f)$ by $\phi(t=\tau;r,f)=e^{-1}$, and then define the ``susceptibility'' of the relaxation time: 
\begin{equation}
\chi(r, f) = \frac{\tau(r,f) - \tau(f=0)}{f} .
\end{equation}
Here, the reference relaxation time $\tau(f=0)$ is taken from the equilibrium results in Sec.~\ref{section_overlap}. 
We compute this susceptibility $\chi(r, f)$ at various temperatures at $f=100$, which is in the linear response regime. 

Fig.~\ref{fig_chi} shows the results at the target temperatures~\footnote{
The error bars in Fig.~\ref{fig_chi} were derived as follows. 
We first calculated the sample mean of $\phi(t;r,f)$ and its standard error $E(t;r,f)$. 
We then obtained $\phi_{\pm}(t;r,f) = \phi(t;r,f) \pm E(t;r,f)$ and computed $\chi_{\pm}(r)$, which defines the upper and lower bounds of our error bars.}. 
At higher temperatures $T= 15$ and 19, the susceptibility $\chi (r, f)$ is negligibly small, meaning that the structural relaxation is insensitive to the external force. 
In contrast, at lower temperatures $T=12$ and 11, $\chi (r, f)$ deviates strongly from 0 and becomes negative for small $r$. 
This means that the structural relaxation becomes faster near the external force in the deeply supercooled liquids. 

The observed behavior of the susceptibility $\chi (r, f)$ is related to dynamic heterogeneity of supercooled liquids.  
It is known that deeper supercooling markedly slows structural relaxation and increases the length scale of dynamic heterogeneity~\cite{Cavagna2009review}. 
Dynamic heterogeneity has been quantified using various techniques~\cite{Lacevic2003, Berthier2004, Toninelli2005, Karmakar2009, Kawasaki2010}, and as one of such technique, Kim et al. studied a two-point density correlation function of supercooled liquids under a sinusoidal external force field $
\vec{f}_{\mathrm{Kim},\vec{q}}(\vec{r}) = i\vec{q}f \exp [ -i\vec{q}\cdot \vec{r}],
$
where $\vec{q}$ is a wavenumber argument and $f$ is the amplitude of the force. 
They measured $\chi_U$, the susceptibility of the two-point density correlation function in this setting, to estimate the length scale of dynamic heterogeneity~\cite{Kim2013}. 
Now, our dipolar force field is related to this sinusoidal external force as 
\begin{equation}
\vec{f} (\vec{r}) \propto \int d\vec{q} \: { \vec{f}_{\mathrm{Kim},\vec{q}}(\vec{r}) \cdot \vec{e}_x } .
\label{eq_kim_our_field}
\end{equation}
This means that $\chi(r, f)$ studied by us is linearly related to $\chi_U$ studied by Kim et al. 
Since  $\chi_U$ captures dynamic heterogeneity, the deviation of $\chi(r, f)$ is quite possibly related to the increase of dynamic heterogeneity. 

In this section, we studied $\chi(r, f)$ to find that the structural relaxation becomes faster near the dipolar force field. 
We discussed that this deviation of $\chi(r, f)$ is related to dynamic heterogeneity. 
The deviation of $\chi(r, f)$ occurs on a similar length scale at which $\frac{I_\mathrm{MD}(r)}{I_\mathrm{NS}(r)}$ also deviates from 1 (see Figs.~\ref{main_fig} and \ref{fig_chi}). 
This suggests a connection between dynamic heterogeneity and the violation of the hydrodynamic description in supercooled liquids.

\section{Summary and Discussion}

This study used MD simulations to study the steady flow of supercooled liquids induced by a localized external force. 
To characterize the system, we first studied the equilibrium dynamics of our model.  
By measuring the mean-squared displacement, overlap function, and stress autocorrelation function, we showed that our model exhibits typical dynamical behaviors of fragile supercooled liquids: e.g., the two-step relaxation and super-Arrhenius increase of the viscosity. 

Then, we performed MD simulations of our model under an external dipolar force field. 
On the one hand, we measured the velocity field induced by the external force using MD simulations. 
On the other hand, we computed the velocity field using the Navier-Stokes equations with the viscosity data obtained in the equilibrium MD simulations. 
By comparing these two results, we revealed that the flow velocity exceeds the hydrodynamic prediction in proximity to the external force field. 
This is true even for small external forces within the linear response regime, and this deviation becomes increasingly pronounced with deeper supercooling. 
Therefore, our results demonstrated the breakdown of the Navier-Stokes equations in the supercooled liquids in real space. 

Finally, we investigated the structural relaxation of the supercooled liquids in the steady flow states. 
We measured the overlap function in the steady flow states and calculated the susceptibility $\chi(r,f)$ of the relaxation time to the external force. 
We found that the structural relaxation in the vicinity of the external force progresses more rapidly in systems subjected to the dipolar external force field. 
Furthermore, this acceleration in relaxation time, paralleling the increase in flow velocity, was more pronounced with further supercooling. 
This correlation implies a connection between dynamic heterogeneity and the violation of the hydrodynamic description.

Our results suggest that a characteristic length scale below which the hydrodynamics breaks down emerges in supercooled liquids. 
Unfortunately, the present MD simulations lack the precision required to quantitatively ascertain this length scale. 
This limitation primarily arises from the huge computational effort required. 
The challenge is compounded by the large thermal fluctuations inherent in liquids, which necessitate extended simulation durations to average out these fluctuations and accurately discern the response.
It is then an interesting future work to perform larger-scale MD simulations to quantify this length scale. 

Furthermore, it is also promising to analyze our system using the Mode-Coupling Theory (MCT) ~\cite{Gotze2008}. 
The MCT is a first-principles-based framework that provides a quantitative description of the dynamic anomalies in supercooled liquids near the glass transition ~\cite{MCT_Janssen}. 
The MCT was originally developed for equilibrium systems but was subsequently extended to systems with an external field, which is called the Inhomogeneous Mode-Coupling Theory (IMCT)~\cite{Biroli2006}. 
The IMCT addresses the behavior of supercooled liquids under perturbation, making it relevant to our setting.
Although the IMCT was formulated to study the system under a sinusoidal external force field, this theory could be extended to the system under the dipolar force field. 
The study along this line could further elucidate the relationship between dynamic heterogeneity and the breakdown of the Navier-Stokes equations. 

\begin{acknowledgments}
We thank H.~Mizuno and N.~Oyama for insightful discussions.
This work was supported by JSPS KAKENHI Grant Numbers JP20H01868 and JP24H00192.
\end{acknowledgments}

\bibliography{citations}

\begin{thebibliography}{50}%
\makeatletter
\providecommand \@ifxundefined [1]{%
 \@ifx{#1\undefined}
}%
\providecommand \@ifnum [1]{%
 \ifnum #1\expandafter \@firstoftwo
 \else \expandafter \@secondoftwo
 \fi
}%
\providecommand \@ifx [1]{%
 \ifx #1\expandafter \@firstoftwo
 \else \expandafter \@secondoftwo
 \fi
}%
\providecommand \natexlab [1]{#1}%
\providecommand \enquote  [1]{``#1''}%
\providecommand \bibnamefont  [1]{#1}%
\providecommand \bibfnamefont [1]{#1}%
\providecommand \citenamefont [1]{#1}%
\providecommand \href@noop [0]{\@secondoftwo}%
\providecommand \href [0]{\begingroup \@sanitize@url \@href}%
\providecommand \@href[1]{\@@startlink{#1}\@@href}%
\providecommand \@@href[1]{\endgroup#1\@@endlink}%
\providecommand \@sanitize@url [0]{\catcode `\\12\catcode `\$12\catcode `\&12\catcode `\#12\catcode `\^12\catcode `\_12\catcode `\%12\relax}%
\providecommand \@@startlink[1]{}%
\providecommand \@@endlink[0]{}%
\providecommand \url  [0]{\begingroup\@sanitize@url \@url }%
\providecommand \@url [1]{\endgroup\@href {#1}{\urlprefix }}%
\providecommand \urlprefix  [0]{URL }%
\providecommand \Eprint [0]{\href }%
\providecommand \doibase [0]{http://dx.doi.org/}%
\providecommand \selectlanguage [0]{\@gobble}%
\providecommand \bibinfo  [0]{\@secondoftwo}%
\providecommand \bibfield  [0]{\@secondoftwo}%
\providecommand \translation [1]{[#1]}%
\providecommand \BibitemOpen [0]{}%
\providecommand \bibitemStop [0]{}%
\providecommand \bibitemNoStop [0]{.\EOS\space}%
\providecommand \EOS [0]{\spacefactor3000\relax}%
\providecommand \BibitemShut  [1]{\csname bibitem#1\endcsname}%
\let\auto@bib@innerbib\@empty
\bibitem [{\citenamefont {Angell}(1988)}]{Angell1988}%
  \BibitemOpen
  \bibfield  {author} {\bibinfo {author} {\bibfnamefont {C.~A.}\ \bibnamefont {Angell}},\ }\bibfield  {title} {\enquote {\bibinfo {title} {Structural instability and relaxation in liquid and glassy phases near the fragile liquid limit},}\ }\href {\doibase https://doi.org/10.1016/0022-3093(88)90133-0} {\bibfield  {journal} {\bibinfo  {journal} {Journal of Non-Crystalline Solids}\ }\textbf {\bibinfo {volume} {102}},\ \bibinfo {pages} {205--221} (\bibinfo {year} {1988})},\ \bibinfo {note} {proceedings of the Ninth University Conference on Glass Science}\BibitemShut {NoStop}%
\bibitem [{\citenamefont {Angell}(1995)}]{Angell1995}%
  \BibitemOpen
  \bibfield  {author} {\bibinfo {author} {\bibfnamefont {C.~A.}\ \bibnamefont {Angell}},\ }\bibfield  {title} {\enquote {\bibinfo {title} {Formation of glasses from liquids and biopolymers},}\ }\href {\doibase 10.1126/science.267.5206.1924} {\bibfield  {journal} {\bibinfo  {journal} {Science}\ }\textbf {\bibinfo {volume} {267}},\ \bibinfo {pages} {1924--1935} (\bibinfo {year} {1995})},\ \Eprint {http://arxiv.org/abs/https://www.science.org/doi/pdf/10.1126/science.267.5206.1924} {https://www.science.org/doi/pdf/10.1126/science.267.5206.1924} \BibitemShut {NoStop}%
\bibitem [{\citenamefont {Ediger}, \citenamefont {Angell},\ and\ \citenamefont {Nagel}(1996)}]{Ediger1996review}%
  \BibitemOpen
  \bibfield  {author} {\bibinfo {author} {\bibfnamefont {M.~D.}\ \bibnamefont {Ediger}}, \bibinfo {author} {\bibfnamefont {C.~A.}\ \bibnamefont {Angell}}, \ and\ \bibinfo {author} {\bibfnamefont {S.~R.}\ \bibnamefont {Nagel}},\ }\bibfield  {title} {\enquote {\bibinfo {title} {Supercooled liquids and glasses},}\ }\href@noop {} {\bibfield  {journal} {\bibinfo  {journal} {J. Phys. Chem.}\ }\textbf {\bibinfo {volume} {100}},\ \bibinfo {pages} {13200} (\bibinfo {year} {1996})}\BibitemShut {NoStop}%
\bibitem [{\citenamefont {Debenedetti}\ and\ \citenamefont {Stillinger}(2001)}]{Debenedetti2001review}%
  \BibitemOpen
  \bibfield  {author} {\bibinfo {author} {\bibfnamefont {P.~G.}\ \bibnamefont {Debenedetti}}\ and\ \bibinfo {author} {\bibfnamefont {F.~H.}\ \bibnamefont {Stillinger}},\ }\bibfield  {title} {\enquote {\bibinfo {title} {Supercooled liquids and the glass transition},}\ }\href@noop {} {\bibfield  {journal} {\bibinfo  {journal} {Nature}\ }\textbf {\bibinfo {volume} {410}},\ \bibinfo {pages} {259} (\bibinfo {year} {2001})}\BibitemShut {NoStop}%
\bibitem [{\citenamefont {Cavagna}(2009)}]{Cavagna2009review}%
  \BibitemOpen
  \bibfield  {author} {\bibinfo {author} {\bibfnamefont {A.}~\bibnamefont {Cavagna}},\ }\bibfield  {title} {\enquote {\bibinfo {title} {Supercooled liquids for pedestrians},}\ }\href@noop {} {\bibfield  {journal} {\bibinfo  {journal} {Phys. Rep.}\ }\textbf {\bibinfo {volume} {476}},\ \bibinfo {pages} {51} (\bibinfo {year} {2009})}\BibitemShut {NoStop}%
\bibitem [{\citenamefont {Gotze}\ and\ \citenamefont {Sjogren}(1992)}]{Gotze1992}%
  \BibitemOpen
  \bibfield  {author} {\bibinfo {author} {\bibfnamefont {W.}~\bibnamefont {Gotze}}\ and\ \bibinfo {author} {\bibfnamefont {L.}~\bibnamefont {Sjogren}},\ }\bibfield  {title} {\enquote {\bibinfo {title} {Relaxation processes in supercooled liquids},}\ }\href {\doibase 10.1088/0034-4885/55/3/001} {\bibfield  {journal} {\bibinfo  {journal} {Reports on Progress in Physics}\ }\textbf {\bibinfo {volume} {55}},\ \bibinfo {pages} {241} (\bibinfo {year} {1992})}\BibitemShut {NoStop}%
\bibitem [{\citenamefont {Yamamoto}\ and\ \citenamefont {Onuki}(1998)}]{Yamamoto-Onuki1998}%
  \BibitemOpen
  \bibfield  {author} {\bibinfo {author} {\bibfnamefont {R.}~\bibnamefont {Yamamoto}}\ and\ \bibinfo {author} {\bibfnamefont {A.}~\bibnamefont {Onuki}},\ }\bibfield  {title} {\enquote {\bibinfo {title} {Dynamics of highly supercooled liquids: Heterogeneity, rheology, and diffusion},}\ }\href@noop {} {\bibfield  {journal} {\bibinfo  {journal} {Phys. Rev. E}\ }\textbf {\bibinfo {volume} {58}},\ \bibinfo {pages} {3515} (\bibinfo {year} {1998})}\BibitemShut {NoStop}%
\bibitem [{\citenamefont {Tong}\ and\ \citenamefont {Tanaka}(2018)}]{Tong2018}%
  \BibitemOpen
  \bibfield  {author} {\bibinfo {author} {\bibfnamefont {H.}~\bibnamefont {Tong}}\ and\ \bibinfo {author} {\bibfnamefont {H.}~\bibnamefont {Tanaka}},\ }\bibfield  {title} {\enquote {\bibinfo {title} {Revealing hidden structural order controlling both fast and slow glassy dynamics in supercooled liquids},}\ }\href {\doibase 10.1103/PhysRevX.8.011041} {\bibfield  {journal} {\bibinfo  {journal} {Phys. Rev. X}\ }\textbf {\bibinfo {volume} {8}},\ \bibinfo {pages} {011041} (\bibinfo {year} {2018})}\BibitemShut {NoStop}%
\bibitem [{\citenamefont {Berthier}\ and\ \citenamefont {Biroli}(2011)}]{Berthier2011review}%
  \BibitemOpen
  \bibfield  {author} {\bibinfo {author} {\bibfnamefont {L.}~\bibnamefont {Berthier}}\ and\ \bibinfo {author} {\bibfnamefont {G.}~\bibnamefont {Biroli}},\ }\bibfield  {title} {\enquote {\bibinfo {title} {Theoretical perspective on the glass transition and amorphous materials},}\ }\href@noop {} {\bibfield  {journal} {\bibinfo  {journal} {Rev. Mod. Phys.}\ }\textbf {\bibinfo {volume} {83}},\ \bibinfo {pages} {587} (\bibinfo {year} {2011})}\BibitemShut {NoStop}%
\bibitem [{\citenamefont {Hurley}\ and\ \citenamefont {Harrowell}(1995)}]{Hurley-Harrowell1995}%
  \BibitemOpen
  \bibfield  {author} {\bibinfo {author} {\bibfnamefont {M.~M.}\ \bibnamefont {Hurley}}\ and\ \bibinfo {author} {\bibfnamefont {P.}~\bibnamefont {Harrowell}},\ }\bibfield  {title} {\enquote {\bibinfo {title} {Kinetic structure of a two-dimensional liquid},}\ }\href@noop {} {\bibfield  {journal} {\bibinfo  {journal} {Phys. Rev. E}\ }\textbf {\bibinfo {volume} {52}},\ \bibinfo {pages} {1694} (\bibinfo {year} {1995})}\BibitemShut {NoStop}%
\bibitem [{\citenamefont {Kob}\ \emph {et~al.}(1997)\citenamefont {Kob}, \citenamefont {Donati}, \citenamefont {Plimpton}, \citenamefont {Poole},\ and\ \citenamefont {Glotzer}}]{Kob1997}%
  \BibitemOpen
  \bibfield  {author} {\bibinfo {author} {\bibfnamefont {W.}~\bibnamefont {Kob}}, \bibinfo {author} {\bibfnamefont {C.}~\bibnamefont {Donati}}, \bibinfo {author} {\bibfnamefont {S.~J.}\ \bibnamefont {Plimpton}}, \bibinfo {author} {\bibfnamefont {P.~H.}\ \bibnamefont {Poole}}, \ and\ \bibinfo {author} {\bibfnamefont {S.~C.}\ \bibnamefont {Glotzer}},\ }\bibfield  {title} {\enquote {\bibinfo {title} {Dynamical heterogeneities in a supercooled lennard-jones liquid},}\ }\href@noop {} {\bibfield  {journal} {\bibinfo  {journal} {Phys. Rev. Lett.}\ }\textbf {\bibinfo {volume} {79}},\ \bibinfo {pages} {2827} (\bibinfo {year} {1997})}\BibitemShut {NoStop}%
\bibitem [{\citenamefont {Yamamoto}\ and\ \citenamefont {Onuki}(1997)}]{Yamamoto-Onuki1997}%
  \BibitemOpen
  \bibfield  {author} {\bibinfo {author} {\bibfnamefont {R.}~\bibnamefont {Yamamoto}}\ and\ \bibinfo {author} {\bibfnamefont {A.}~\bibnamefont {Onuki}},\ }\bibfield  {title} {\enquote {\bibinfo {title} {Kinetic heterogeneities in a highly supercooled liquid},}\ }\href@noop {} {\bibfield  {journal} {\bibinfo  {journal} {J. Phys. Soc. Jpn.}\ }\textbf {\bibinfo {volume} {66}},\ \bibinfo {pages} {2545} (\bibinfo {year} {1997})}\BibitemShut {NoStop}%
\bibitem [{\citenamefont {Donati}\ \emph {et~al.}(1998)\citenamefont {Donati}, \citenamefont {Douglas}, \citenamefont {Kob}, \citenamefont {Plimpton}, \citenamefont {Poole},\ and\ \citenamefont {Glotzer}}]{Donati1998}%
  \BibitemOpen
  \bibfield  {author} {\bibinfo {author} {\bibfnamefont {C.}~\bibnamefont {Donati}}, \bibinfo {author} {\bibfnamefont {J.~F.}\ \bibnamefont {Douglas}}, \bibinfo {author} {\bibfnamefont {W.}~\bibnamefont {Kob}}, \bibinfo {author} {\bibfnamefont {S.~J.}\ \bibnamefont {Plimpton}}, \bibinfo {author} {\bibfnamefont {P.~H.}\ \bibnamefont {Poole}}, \ and\ \bibinfo {author} {\bibfnamefont {S.~C.}\ \bibnamefont {Glotzer}},\ }\bibfield  {title} {\enquote {\bibinfo {title} {Stringlike cooperative motion in a supercooled liquid},}\ }\href {\doibase 10.1103/PhysRevLett.80.2338} {\bibfield  {journal} {\bibinfo  {journal} {Phys. Rev. Lett.}\ }\textbf {\bibinfo {volume} {80}},\ \bibinfo {pages} {2338--2341} (\bibinfo {year} {1998})}\BibitemShut {NoStop}%
\bibitem [{\citenamefont {Ediger}(2000)}]{Ediger2000review}%
  \BibitemOpen
  \bibfield  {author} {\bibinfo {author} {\bibfnamefont {M.~D.}\ \bibnamefont {Ediger}},\ }\bibfield  {title} {\enquote {\bibinfo {title} {Spatially heterogeneous dynamics in supercooled liquids},}\ }\href {\doibase 10.1146/annurev.physchem.51.1.99} {\bibfield  {journal} {\bibinfo  {journal} {Annual Review of Physical Chemistry}\ }\textbf {\bibinfo {volume} {51}},\ \bibinfo {pages} {99--128} (\bibinfo {year} {2000})},\ \bibinfo {note} {pMID: 11031277},\ \Eprint {http://arxiv.org/abs/https://doi.org/10.1146/annurev.physchem.51.1.99} {https://doi.org/10.1146/annurev.physchem.51.1.99} \BibitemShut {NoStop}%
\bibitem [{\citenamefont {Weeks}\ \emph {et~al.}(2000)\citenamefont {Weeks}, \citenamefont {Crocker}, \citenamefont {Levitt}, \citenamefont {Schofield},\ and\ \citenamefont {Weitz}}]{Weeks2000}%
  \BibitemOpen
  \bibfield  {author} {\bibinfo {author} {\bibfnamefont {E.~R.}\ \bibnamefont {Weeks}}, \bibinfo {author} {\bibfnamefont {J.~C.}\ \bibnamefont {Crocker}}, \bibinfo {author} {\bibfnamefont {A.~C.}\ \bibnamefont {Levitt}}, \bibinfo {author} {\bibfnamefont {A.}~\bibnamefont {Schofield}}, \ and\ \bibinfo {author} {\bibfnamefont {D.~A.}\ \bibnamefont {Weitz}},\ }\bibfield  {title} {\enquote {\bibinfo {title} {Three-dimensional direct imaging of structural relaxation near the colloidal glass transition},}\ }\href {\doibase 10.1126/science.287.5453.627} {\bibfield  {journal} {\bibinfo  {journal} {Science}\ }\textbf {\bibinfo {volume} {287}},\ \bibinfo {pages} {627--631} (\bibinfo {year} {2000})},\ \Eprint {http://arxiv.org/abs/https://www.science.org/doi/pdf/10.1126/science.287.5453.627} {https://www.science.org/doi/pdf/10.1126/science.287.5453.627} \BibitemShut {NoStop}%
\bibitem [{\citenamefont {Lačević}\ \emph {et~al.}(2002)\citenamefont {Lačević}, \citenamefont {Starr}, \citenamefont {Schrøder}, \citenamefont {Novikov},\ and\ \citenamefont {Glotzer}}]{Lacevic2002}%
  \BibitemOpen
  \bibfield  {author} {\bibinfo {author} {\bibfnamefont {N.}~\bibnamefont {Lačević}}, \bibinfo {author} {\bibfnamefont {F.}~\bibnamefont {Starr}}, \bibinfo {author} {\bibfnamefont {T.}~\bibnamefont {Schrøder}}, \bibinfo {author} {\bibfnamefont {V.}~\bibnamefont {Novikov}}, \ and\ \bibinfo {author} {\bibfnamefont {S.}~\bibnamefont {Glotzer}},\ }\bibfield  {title} {\enquote {\bibinfo {title} {Growing correlation length on cooling below the onset of caging in a simulated glass-forming liquid},}\ }\href@noop {} {\bibfield  {journal} {\bibinfo  {journal} {Phys. Rev. E}\ }\textbf {\bibinfo {volume} {66}},\ \bibinfo {pages} {030101} (\bibinfo {year} {2002})}\BibitemShut {NoStop}%
\bibitem [{\citenamefont {Lačević}\ \emph {et~al.}(2003)\citenamefont {Lačević}, \citenamefont {Starr}, \citenamefont {Schrøder},\ and\ \citenamefont {Glotzer}}]{Lacevic2003}%
  \BibitemOpen
  \bibfield  {author} {\bibinfo {author} {\bibfnamefont {N.}~\bibnamefont {Lačević}}, \bibinfo {author} {\bibfnamefont {F.~W.}\ \bibnamefont {Starr}}, \bibinfo {author} {\bibfnamefont {T.~B.}\ \bibnamefont {Schrøder}}, \ and\ \bibinfo {author} {\bibfnamefont {S.~C.}\ \bibnamefont {Glotzer}},\ }\bibfield  {title} {\enquote {\bibinfo {title} {Spatially heterogeneous dynamics investigated via a time-dependent four-point density correlation function},}\ }\href@noop {} {\bibfield  {journal} {\bibinfo  {journal} {J. Chem. Phys.}\ }\textbf {\bibinfo {volume} {119}},\ \bibinfo {pages} {7372} (\bibinfo {year} {2003})}\BibitemShut {NoStop}%
\bibitem [{\citenamefont {Widmer-Cooper}, \citenamefont {Harrowell},\ and\ \citenamefont {Fynewever}(2004)}]{Widmer2004}%
  \BibitemOpen
  \bibfield  {author} {\bibinfo {author} {\bibfnamefont {A.}~\bibnamefont {Widmer-Cooper}}, \bibinfo {author} {\bibfnamefont {P.}~\bibnamefont {Harrowell}}, \ and\ \bibinfo {author} {\bibfnamefont {H.}~\bibnamefont {Fynewever}},\ }\bibfield  {title} {\enquote {\bibinfo {title} {How reproducible are dynamic heterogeneities in a supercooled liquid?}}\ }\href {\doibase 10.1103/PhysRevLett.93.135701} {\bibfield  {journal} {\bibinfo  {journal} {Phys. Rev. Lett.}\ }\textbf {\bibinfo {volume} {93}},\ \bibinfo {pages} {135701} (\bibinfo {year} {2004})}\BibitemShut {NoStop}%
\bibitem [{\citenamefont {Berthier}(2005)}]{Berthier2005}%
  \BibitemOpen
  \bibfield  {author} {\bibinfo {author} {\bibfnamefont {L.}~\bibnamefont {Berthier}},\ }\bibfield  {title} {\enquote {\bibinfo {title} {Direct experimental evidence of a growing length scale accompanying the glass transition},}\ }\href@noop {} {\bibfield  {journal} {\bibinfo  {journal} {Science}\ }\textbf {\bibinfo {volume} {310}},\ \bibinfo {pages} {1797} (\bibinfo {year} {2005})}\BibitemShut {NoStop}%
\bibitem [{\citenamefont {Dalle-Ferrier}(2007)}]{Dalle-Ferrier2007}%
  \BibitemOpen
  \bibfield  {author} {\bibinfo {author} {\bibfnamefont {C.}~\bibnamefont {Dalle-Ferrier}},\ }\bibfield  {title} {\enquote {\bibinfo {title} {Spatial correlations in the dynamics of glassforming liquids: Experimental determination of their temperature dependence},}\ }\href@noop {} {\bibfield  {journal} {\bibinfo  {journal} {Phys. Rev. E}\ }\textbf {\bibinfo {volume} {76}},\ \bibinfo {pages} {041510} (\bibinfo {year} {2007})}\BibitemShut {NoStop}%
\bibitem [{\citenamefont {Flenner}\ and\ \citenamefont {Szamel}(2010)}]{Flenner-Szamel2010}%
  \BibitemOpen
  \bibfield  {author} {\bibinfo {author} {\bibfnamefont {E.}~\bibnamefont {Flenner}}\ and\ \bibinfo {author} {\bibfnamefont {G.}~\bibnamefont {Szamel}},\ }\bibfield  {title} {\enquote {\bibinfo {title} {Dynamic heterogeneity in a glass forming fluid: Susceptibility, structure factor, and correlation length},}\ }\href@noop {} {\bibfield  {journal} {\bibinfo  {journal} {Phys. Rev. Lett.}\ }\textbf {\bibinfo {volume} {105}},\ \bibinfo {pages} {217801} (\bibinfo {year} {2010})}\BibitemShut {NoStop}%
\bibitem [{\citenamefont {Flenner}, \citenamefont {Zhang},\ and\ \citenamefont {Szamel}(2011)}]{Flenner2011}%
  \BibitemOpen
  \bibfield  {author} {\bibinfo {author} {\bibfnamefont {E.}~\bibnamefont {Flenner}}, \bibinfo {author} {\bibfnamefont {M.}~\bibnamefont {Zhang}}, \ and\ \bibinfo {author} {\bibfnamefont {G.}~\bibnamefont {Szamel}},\ }\bibfield  {title} {\enquote {\bibinfo {title} {Analysis of a growing dynamic length scale in a glass-forming binary hard-sphere mixture},}\ }\href@noop {} {\bibfield  {journal} {\bibinfo  {journal} {Phys. Rev. E}\ }\textbf {\bibinfo {volume} {83}},\ \bibinfo {pages} {051501} (\bibinfo {year} {2011})}\BibitemShut {NoStop}%
\bibitem [{\citenamefont {Shiba}, \citenamefont {Kawasaki},\ and\ \citenamefont {Onuki}(2012)}]{Shiba2012}%
  \BibitemOpen
  \bibfield  {author} {\bibinfo {author} {\bibfnamefont {H.}~\bibnamefont {Shiba}}, \bibinfo {author} {\bibfnamefont {T.}~\bibnamefont {Kawasaki}}, \ and\ \bibinfo {author} {\bibfnamefont {A.}~\bibnamefont {Onuki}},\ }\bibfield  {title} {\enquote {\bibinfo {title} {Relationship between bond-breakage correlations and four-point correlations in heterogeneous glassy dynamics: Configuration changes and vibration modes},}\ }\href {\doibase 10.1103/PhysRevE.86.041504} {\bibfield  {journal} {\bibinfo  {journal} {Phys. Rev. E}\ }\textbf {\bibinfo {volume} {86}},\ \bibinfo {pages} {041504} (\bibinfo {year} {2012})}\BibitemShut {NoStop}%
\bibitem [{\citenamefont {Flenner}, \citenamefont {Staley},\ and\ \citenamefont {Szamel}(2014)}]{Flenner2014}%
  \BibitemOpen
  \bibfield  {author} {\bibinfo {author} {\bibfnamefont {E.}~\bibnamefont {Flenner}}, \bibinfo {author} {\bibfnamefont {H.}~\bibnamefont {Staley}}, \ and\ \bibinfo {author} {\bibfnamefont {G.}~\bibnamefont {Szamel}},\ }\bibfield  {title} {\enquote {\bibinfo {title} {Universal features of dynamic heterogeneity in supercooled liquids},}\ }\href {\doibase 10.1103/PhysRevLett.112.097801} {\bibfield  {journal} {\bibinfo  {journal} {Phys. Rev. Lett.}\ }\textbf {\bibinfo {volume} {112}},\ \bibinfo {pages} {097801} (\bibinfo {year} {2014})}\BibitemShut {NoStop}%
\bibitem [{\citenamefont {Hansen}\ and\ \citenamefont {McDonald}(2013)}]{Hansen2013}%
  \BibitemOpen
  \bibfield  {author} {\bibinfo {author} {\bibfnamefont {J.-P.}\ \bibnamefont {Hansen}}\ and\ \bibinfo {author} {\bibfnamefont {I.~R.}\ \bibnamefont {McDonald}},\ }\href {\doibase https://doi.org/10.1016/B978-0-12-387032-2.00013-1} {\emph {\bibinfo {title} {Theory of Simple Liquids (Fourth Edition)}}},\ \bibinfo {edition} {fourth edition}\ ed.\ (\bibinfo  {publisher} {Academic Press},\ \bibinfo {address} {Oxford},\ \bibinfo {year} {2013})\BibitemShut {NoStop}%
\bibitem [{\citenamefont {Kim}\ and\ \citenamefont {Karrila}(2005)}]{microhydrodynamics}%
  \BibitemOpen
  \bibfield  {author} {\bibinfo {author} {\bibfnamefont {S.}~\bibnamefont {Kim}}\ and\ \bibinfo {author} {\bibfnamefont {S.~J.}\ \bibnamefont {Karrila}},\ }\href@noop {} {\emph {\bibinfo {title} {Microhydrodynamics: Principles and Selected Applications}}}\ (\bibinfo  {publisher} {Dover Publications, Inc.},\ \bibinfo {address} {New York},\ \bibinfo {year} {2005})\BibitemShut {NoStop}%
\bibitem [{\citenamefont {Furukawa}\ and\ \citenamefont {Tanaka}(2011)}]{Furukawa2011}%
  \BibitemOpen
  \bibfield  {author} {\bibinfo {author} {\bibfnamefont {A.}~\bibnamefont {Furukawa}}\ and\ \bibinfo {author} {\bibfnamefont {H.}~\bibnamefont {Tanaka}},\ }\bibfield  {title} {\enquote {\bibinfo {title} {Direct evidence of heterogeneous mechanical relaxation in supercooled liquids},}\ }\href {\doibase 10.1103/PhysRevE.84.061503} {\bibfield  {journal} {\bibinfo  {journal} {Phys. Rev. E}\ }\textbf {\bibinfo {volume} {84}},\ \bibinfo {pages} {061503} (\bibinfo {year} {2011})}\BibitemShut {NoStop}%
\bibitem [{\citenamefont {Lerner}\ \emph {et~al.}(2014)\citenamefont {Lerner}, \citenamefont {DeGiuli}, \citenamefont {Düring},\ and\ \citenamefont {Wyart}}]{Lerner2014}%
  \BibitemOpen
  \bibfield  {author} {\bibinfo {author} {\bibfnamefont {E.}~\bibnamefont {Lerner}}, \bibinfo {author} {\bibfnamefont {E.}~\bibnamefont {DeGiuli}}, \bibinfo {author} {\bibfnamefont {G.}~\bibnamefont {Düring}}, \ and\ \bibinfo {author} {\bibfnamefont {M.}~\bibnamefont {Wyart}},\ }\bibfield  {title} {\enquote {\bibinfo {title} {Breakdown of continuum elasticity in amorphous solids},}\ }\href {\doibase 10.1039/C4SM00311J} {\bibfield  {journal} {\bibinfo  {journal} {Soft Matter}\ }\textbf {\bibinfo {volume} {10}},\ \bibinfo {pages} {5085--5092} (\bibinfo {year} {2014})}\BibitemShut {NoStop}%
\bibitem [{\citenamefont {Orts}\ \emph {et~al.}(2020)\citenamefont {Orts}, \citenamefont {Ortega}, \citenamefont {Garz\'on}, \citenamefont {Fuchs},\ and\ \citenamefont {Puertas}}]{exfield-hydro2020}%
  \BibitemOpen
  \bibfield  {author} {\bibinfo {author} {\bibfnamefont {F.}~\bibnamefont {Orts}}, \bibinfo {author} {\bibfnamefont {G.}~\bibnamefont {Ortega}}, \bibinfo {author} {\bibfnamefont {E.~M.}\ \bibnamefont {Garz\'on}}, \bibinfo {author} {\bibfnamefont {M.}~\bibnamefont {Fuchs}}, \ and\ \bibinfo {author} {\bibfnamefont {A.~M.}\ \bibnamefont {Puertas}},\ }\bibfield  {title} {\enquote {\bibinfo {title} {Dynamics and friction of a large colloidal particle in a bath of hard spheres: Langevin dynamics simulations and hydrodynamic description},}\ }\href {\doibase 10.1103/PhysRevE.101.052607} {\bibfield  {journal} {\bibinfo  {journal} {Phys. Rev. E}\ }\textbf {\bibinfo {volume} {101}},\ \bibinfo {pages} {052607} (\bibinfo {year} {2020})}\BibitemShut {NoStop}%
\bibitem [{\citenamefont {Weeks}, \citenamefont {Chandler},\ and\ \citenamefont {Andersen}(2003)}]{WCA2003}%
  \BibitemOpen
  \bibfield  {author} {\bibinfo {author} {\bibfnamefont {J.~D.}\ \bibnamefont {Weeks}}, \bibinfo {author} {\bibfnamefont {D.}~\bibnamefont {Chandler}}, \ and\ \bibinfo {author} {\bibfnamefont {H.~C.}\ \bibnamefont {Andersen}},\ }\bibfield  {title} {\enquote {\bibinfo {title} {{Role of Repulsive Forces in Determining the Equilibrium Structure of Simple Liquids}},}\ }\href {\doibase 10.1063/1.1674820} {\bibfield  {journal} {\bibinfo  {journal} {The Journal of Chemical Physics}\ }\textbf {\bibinfo {volume} {54}},\ \bibinfo {pages} {5237--5247} (\bibinfo {year} {2003})},\ \Eprint {http://arxiv.org/abs/https://pubs.aip.org/aip/jcp/article-pdf/54/12/5237/10967422/5237\_1\_online.pdf} {https://pubs.aip.org/aip/jcp/article-pdf/54/12/5237/10967422/5237\_1\_online.pdf} \BibitemShut {NoStop}%
\bibitem [{\citenamefont {Bussi}, \citenamefont {Donadio},\ and\ \citenamefont {Parrinello}(2007)}]{CSVR2007}%
  \BibitemOpen
  \bibfield  {author} {\bibinfo {author} {\bibfnamefont {G.}~\bibnamefont {Bussi}}, \bibinfo {author} {\bibfnamefont {D.}~\bibnamefont {Donadio}}, \ and\ \bibinfo {author} {\bibfnamefont {M.}~\bibnamefont {Parrinello}},\ }\bibfield  {title} {\enquote {\bibinfo {title} {{Canonical sampling through velocity rescaling}},}\ }\href {\doibase 10.1063/1.2408420} {\bibfield  {journal} {\bibinfo  {journal} {The Journal of Chemical Physics}\ }\textbf {\bibinfo {volume} {126}},\ \bibinfo {pages} {014101} (\bibinfo {year} {2007})},\ \Eprint {http://arxiv.org/abs/https://pubs.aip.org/aip/jcp/article-pdf/doi/10.1063/1.2408420/15393604/014101\_1\_online.pdf} {https://pubs.aip.org/aip/jcp/article-pdf/doi/10.1063/1.2408420/15393604/014101\_1\_online.pdf} \BibitemShut {NoStop}%
\bibitem [{\citenamefont {Shiba}\ \emph {et~al.}(2016)\citenamefont {Shiba}, \citenamefont {Yamada}, \citenamefont {Kawasaki},\ and\ \citenamefont {Kim}}]{Shiba2016}%
  \BibitemOpen
  \bibfield  {author} {\bibinfo {author} {\bibfnamefont {H.}~\bibnamefont {Shiba}}, \bibinfo {author} {\bibfnamefont {Y.}~\bibnamefont {Yamada}}, \bibinfo {author} {\bibfnamefont {T.}~\bibnamefont {Kawasaki}}, \ and\ \bibinfo {author} {\bibfnamefont {K.}~\bibnamefont {Kim}},\ }\bibfield  {title} {\enquote {\bibinfo {title} {Unveiling dimensionality dependence of glassy dynamics: 2d infinite fluctuation eclipses inherent structural relaxation},}\ }\href {\doibase 10.1103/PhysRevLett.117.245701} {\bibfield  {journal} {\bibinfo  {journal} {Phys. Rev. Lett.}\ }\textbf {\bibinfo {volume} {117}},\ \bibinfo {pages} {245701} (\bibinfo {year} {2016})}\BibitemShut {NoStop}%
\bibitem [{\citenamefont {Illing}\ \emph {et~al.}(2017)\citenamefont {Illing}, \citenamefont {Fritschi}, \citenamefont {Kaiser}, \citenamefont {Klix}, \citenamefont {Maret},\ and\ \citenamefont {Keim}}]{Bernd2017}%
  \BibitemOpen
  \bibfield  {author} {\bibinfo {author} {\bibfnamefont {B.}~\bibnamefont {Illing}}, \bibinfo {author} {\bibfnamefont {S.}~\bibnamefont {Fritschi}}, \bibinfo {author} {\bibfnamefont {H.}~\bibnamefont {Kaiser}}, \bibinfo {author} {\bibfnamefont {C.~L.}\ \bibnamefont {Klix}}, \bibinfo {author} {\bibfnamefont {G.}~\bibnamefont {Maret}}, \ and\ \bibinfo {author} {\bibfnamefont {P.}~\bibnamefont {Keim}},\ }\bibfield  {title} {\enquote {\bibinfo {title} {Mermin–wagner fluctuations in 2d amorphous solids},}\ }\href {\doibase 10.1073/pnas.1612964114} {\bibfield  {journal} {\bibinfo  {journal} {Proceedings of the National Academy of Sciences}\ }\textbf {\bibinfo {volume} {114}},\ \bibinfo {pages} {1856--1861} (\bibinfo {year} {2017})},\ \Eprint {http://arxiv.org/abs/https://www.pnas.org/doi/pdf/10.1073/pnas.1612964114} {https://www.pnas.org/doi/pdf/10.1073/pnas.1612964114} \BibitemShut {NoStop}%
\bibitem [{\citenamefont {Flenner}\ and\ \citenamefont {Szamel}(2019)}]{Flenner-Szamel2019}%
  \BibitemOpen
  \bibfield  {author} {\bibinfo {author} {\bibfnamefont {E.}~\bibnamefont {Flenner}}\ and\ \bibinfo {author} {\bibfnamefont {G.}~\bibnamefont {Szamel}},\ }\bibfield  {title} {\enquote {\bibinfo {title} {Viscoelastic shear stress relaxation in two-dimensional glass-forming liquids},}\ }\href {\doibase 10.1073/pnas.1815097116} {\bibfield  {journal} {\bibinfo  {journal} {Proceedings of the National Academy of Sciences}\ }\textbf {\bibinfo {volume} {116}},\ \bibinfo {pages} {2015--2020} (\bibinfo {year} {2019})},\ \Eprint {http://arxiv.org/abs/https://www.pnas.org/doi/pdf/10.1073/pnas.1815097116} {https://www.pnas.org/doi/pdf/10.1073/pnas.1815097116} \BibitemShut {NoStop}%
\bibitem [{\citenamefont {Li}\ \emph {et~al.}(2019)\citenamefont {Li}, \citenamefont {Mishra}, \citenamefont {Sun}, \citenamefont {Zhao}, \citenamefont {Mason}, \citenamefont {Ganapathy},\ and\ \citenamefont {Ciamarra}}]{Li2019}%
  \BibitemOpen
  \bibfield  {author} {\bibinfo {author} {\bibfnamefont {Y.-W.}\ \bibnamefont {Li}}, \bibinfo {author} {\bibfnamefont {C.~K.}\ \bibnamefont {Mishra}}, \bibinfo {author} {\bibfnamefont {Z.-Y.}\ \bibnamefont {Sun}}, \bibinfo {author} {\bibfnamefont {K.}~\bibnamefont {Zhao}}, \bibinfo {author} {\bibfnamefont {T.~G.}\ \bibnamefont {Mason}}, \bibinfo {author} {\bibfnamefont {R.}~\bibnamefont {Ganapathy}}, \ and\ \bibinfo {author} {\bibfnamefont {M.~P.}\ \bibnamefont {Ciamarra}},\ }\bibfield  {title} {\enquote {\bibinfo {title} {Long-wavelength fluctuations and anomalous dynamics in 2-dimensional liquids},}\ }\href {\doibase 10.1073/pnas.1909319116} {\bibfield  {journal} {\bibinfo  {journal} {Proceedings of the National Academy of Sciences}\ }\textbf {\bibinfo {volume} {116}},\ \bibinfo {pages} {22977--22982} (\bibinfo {year} {2019})},\ \Eprint {http://arxiv.org/abs/https://www.pnas.org/doi/pdf/10.1073/pnas.1909319116} {https://www.pnas.org/doi/pdf/10.1073/pnas.1909319116} \BibitemShut {NoStop}%
\bibitem [{\citenamefont {Doliwa}\ and\ \citenamefont {Heuer}(1998)}]{Cage-Effect1998}%
  \BibitemOpen
  \bibfield  {author} {\bibinfo {author} {\bibfnamefont {B.}~\bibnamefont {Doliwa}}\ and\ \bibinfo {author} {\bibfnamefont {A.}~\bibnamefont {Heuer}},\ }\bibfield  {title} {\enquote {\bibinfo {title} {Cage effect, local anisotropies, and dynamic heterogeneities at the glass transition: A computer study of hard spheres},}\ }\href {\doibase 10.1103/PhysRevLett.80.4915} {\bibfield  {journal} {\bibinfo  {journal} {Phys. Rev. Lett.}\ }\textbf {\bibinfo {volume} {80}},\ \bibinfo {pages} {4915--4918} (\bibinfo {year} {1998})}\BibitemShut {NoStop}%
\bibitem [{\citenamefont {Shiba}, \citenamefont {Keim},\ and\ \citenamefont {Kawasaki}(2018)}]{Shiba2018}%
  \BibitemOpen
  \bibfield  {author} {\bibinfo {author} {\bibfnamefont {H.}~\bibnamefont {Shiba}}, \bibinfo {author} {\bibfnamefont {P.}~\bibnamefont {Keim}}, \ and\ \bibinfo {author} {\bibfnamefont {T.}~\bibnamefont {Kawasaki}},\ }\bibfield  {title} {\enquote {\bibinfo {title} {Isolating long-wavelength fluctuation from structural relaxation in two-dimensional glass: cage-relative displacement},}\ }\href {\doibase 10.1088/1361-648X/aaa8b8} {\bibfield  {journal} {\bibinfo  {journal} {Journal of Physics: Condensed Matter}\ }\textbf {\bibinfo {volume} {30}},\ \bibinfo {pages} {094004} (\bibinfo {year} {2018})}\BibitemShut {NoStop}%
\bibitem [{\citenamefont {Varnik}(2006)}]{Varnik2006}%
  \BibitemOpen
  \bibfield  {author} {\bibinfo {author} {\bibfnamefont {F.}~\bibnamefont {Varnik}},\ }\bibfield  {title} {\enquote {\bibinfo {title} {{Structural relaxation and rheological response of a driven amorphous system}},}\ }\href {\doibase 10.1063/1.2363998} {\bibfield  {journal} {\bibinfo  {journal} {The Journal of Chemical Physics}\ }\textbf {\bibinfo {volume} {125}},\ \bibinfo {pages} {164514} (\bibinfo {year} {2006})},\ \Eprint {http://arxiv.org/abs/https://pubs.aip.org/aip/jcp/article-pdf/doi/10.1063/1.2363998/13712520/164514\_1\_online.pdf} {https://pubs.aip.org/aip/jcp/article-pdf/doi/10.1063/1.2363998/13712520/164514\_1\_online.pdf} \BibitemShut {NoStop}%
\bibitem [{\citenamefont {Kim}\ and\ \citenamefont {Keyes}(2005)}]{k-dependent-visc-Kim2005}%
  \BibitemOpen
  \bibfield  {author} {\bibinfo {author} {\bibfnamefont {J.}~\bibnamefont {Kim}}\ and\ \bibinfo {author} {\bibfnamefont {T.}~\bibnamefont {Keyes}},\ }\bibfield  {title} {\enquote {\bibinfo {title} {On the breakdown of the stokes-einstein law in supercooled liquids},}\ }\href {\doibase 10.1021/jp052338r} {\bibfield  {journal} {\bibinfo  {journal} {The Journal of Physical Chemistry B}\ }\textbf {\bibinfo {volume} {109}},\ \bibinfo {pages} {21445--21448} (\bibinfo {year} {2005})}\BibitemShut {NoStop}%
\bibitem [{\citenamefont {Furukawa}\ and\ \citenamefont {Tanaka}(2009)}]{k-dependent-visc-Furukawa2009}%
  \BibitemOpen
  \bibfield  {author} {\bibinfo {author} {\bibfnamefont {A.}~\bibnamefont {Furukawa}}\ and\ \bibinfo {author} {\bibfnamefont {H.}~\bibnamefont {Tanaka}},\ }\bibfield  {title} {\enquote {\bibinfo {title} {Nonlocal nature of the viscous transport in supercooled liquids: Complex fluid approach to supercooled liquids},}\ }\href {\doibase 10.1103/PhysRevLett.103.135703} {\bibfield  {journal} {\bibinfo  {journal} {Phys. Rev. Lett.}\ }\textbf {\bibinfo {volume} {103}},\ \bibinfo {pages} {135703} (\bibinfo {year} {2009})}\BibitemShut {NoStop}%
\bibitem [{\citenamefont {Puscasu}\ \emph {et~al.}(2010)\citenamefont {Puscasu}, \citenamefont {Todd}, \citenamefont {Daivis},\ and\ \citenamefont {Hansen}}]{k-dependent-visc-Puscasu2010}%
  \BibitemOpen
  \bibfield  {author} {\bibinfo {author} {\bibfnamefont {R.~M.}\ \bibnamefont {Puscasu}}, \bibinfo {author} {\bibfnamefont {B.~D.}\ \bibnamefont {Todd}}, \bibinfo {author} {\bibfnamefont {P.~J.}\ \bibnamefont {Daivis}}, \ and\ \bibinfo {author} {\bibfnamefont {J.~S.}\ \bibnamefont {Hansen}},\ }\bibfield  {title} {\enquote {\bibinfo {title} {{Nonlocal viscosity of polymer melts approaching their glassy state}},}\ }\href {\doibase 10.1063/1.3499745} {\bibfield  {journal} {\bibinfo  {journal} {The Journal of Chemical Physics}\ }\textbf {\bibinfo {volume} {133}},\ \bibinfo {pages} {144907} (\bibinfo {year} {2010})},\ \Eprint {http://arxiv.org/abs/https://pubs.aip.org/aip/jcp/article-pdf/doi/10.1063/1.3499745/15431445/144907\_1\_online.pdf} {https://pubs.aip.org/aip/jcp/article-pdf/doi/10.1063/1.3499745/15431445/144907\_1\_online.pdf} \BibitemShut {NoStop}%
\bibitem [{Note1()}]{Note1}%
  \BibitemOpen
  \bibinfo {note} {The error bars in Fig.~\ref {fig_chi} were derived as follows. We first calculated the sample mean of $\phi (t;r,f)$ and its standard error $E(t;r,f)$. We then obtained $\phi _{\pm }(t;r,f) = \phi (t;r,f) \pm E(t;r,f)$ and computed $\chi _{\pm }(r)$, which defines the upper and lower bounds of our error bars.}\BibitemShut {Stop}%
\bibitem [{\citenamefont {Berthier}(2004)}]{Berthier2004}%
  \BibitemOpen
  \bibfield  {author} {\bibinfo {author} {\bibfnamefont {L.}~\bibnamefont {Berthier}},\ }\bibfield  {title} {\enquote {\bibinfo {title} {Time and length scales in supercooled liquids},}\ }\href@noop {} {\bibfield  {journal} {\bibinfo  {journal} {Phys. Rev. E}\ }\textbf {\bibinfo {volume} {69}},\ \bibinfo {pages} {020201R} (\bibinfo {year} {2004})}\BibitemShut {NoStop}%
\bibitem [{\citenamefont {Toninelli}\ \emph {et~al.}(2005)\citenamefont {Toninelli}, \citenamefont {Wyart}, \citenamefont {Berthier}, \citenamefont {Biroli},\ and\ \citenamefont {Bouchaud}}]{Toninelli2005}%
  \BibitemOpen
  \bibfield  {author} {\bibinfo {author} {\bibfnamefont {C.}~\bibnamefont {Toninelli}}, \bibinfo {author} {\bibfnamefont {M.}~\bibnamefont {Wyart}}, \bibinfo {author} {\bibfnamefont {L.}~\bibnamefont {Berthier}}, \bibinfo {author} {\bibfnamefont {G.}~\bibnamefont {Biroli}}, \ and\ \bibinfo {author} {\bibfnamefont {J.-P.}\ \bibnamefont {Bouchaud}},\ }\bibfield  {title} {\enquote {\bibinfo {title} {Dynamical susceptibility of glass formers: Contrasting the predictions of theoretical scenarios},}\ }\href {\doibase 10.1103/PhysRevE.71.041505} {\bibfield  {journal} {\bibinfo  {journal} {Phys. Rev. E}\ }\textbf {\bibinfo {volume} {71}},\ \bibinfo {pages} {041505} (\bibinfo {year} {2005})}\BibitemShut {NoStop}%
\bibitem [{\citenamefont {Karmakar}, \citenamefont {Dasgupta},\ and\ \citenamefont {Sastry}(2009)}]{Karmakar2009}%
  \BibitemOpen
  \bibfield  {author} {\bibinfo {author} {\bibfnamefont {S.}~\bibnamefont {Karmakar}}, \bibinfo {author} {\bibfnamefont {C.}~\bibnamefont {Dasgupta}}, \ and\ \bibinfo {author} {\bibfnamefont {S.}~\bibnamefont {Sastry}},\ }\bibfield  {title} {\enquote {\bibinfo {title} {Growing length and time scales in glass-forming liquids},}\ }\href {\doibase 10.1073/pnas.0811082106} {\bibfield  {journal} {\bibinfo  {journal} {Proceedings of the National Academy of Sciences}\ }\textbf {\bibinfo {volume} {106}},\ \bibinfo {pages} {3675--3679} (\bibinfo {year} {2009})},\ \Eprint {http://arxiv.org/abs/https://www.pnas.org/doi/pdf/10.1073/pnas.0811082106} {https://www.pnas.org/doi/pdf/10.1073/pnas.0811082106} \BibitemShut {NoStop}%
\bibitem [{\citenamefont {Kawasaki}\ and\ \citenamefont {Tanaka}(2010)}]{Kawasaki2010}%
  \BibitemOpen
  \bibfield  {author} {\bibinfo {author} {\bibfnamefont {T.}~\bibnamefont {Kawasaki}}\ and\ \bibinfo {author} {\bibfnamefont {H.}~\bibnamefont {Tanaka}},\ }\bibfield  {title} {\enquote {\bibinfo {title} {Structural origin of dynamic heterogeneity in three-dimensional colloidal glass formers and its link to crystal nucleation},}\ }\href {\doibase 10.1088/0953-8984/22/23/232102} {\bibfield  {journal} {\bibinfo  {journal} {Journal of Physics: Condensed Matter}\ }\textbf {\bibinfo {volume} {22}},\ \bibinfo {pages} {232102} (\bibinfo {year} {2010})}\BibitemShut {NoStop}%
\bibitem [{\citenamefont {Kim}\ \emph {et~al.}(2013)\citenamefont {Kim}, \citenamefont {Saito}, \citenamefont {Miyazaki}, \citenamefont {Biroli},\ and\ \citenamefont {Reichman}}]{Kim2013}%
  \BibitemOpen
  \bibfield  {author} {\bibinfo {author} {\bibfnamefont {K.}~\bibnamefont {Kim}}, \bibinfo {author} {\bibfnamefont {S.}~\bibnamefont {Saito}}, \bibinfo {author} {\bibfnamefont {K.}~\bibnamefont {Miyazaki}}, \bibinfo {author} {\bibfnamefont {G.}~\bibnamefont {Biroli}}, \ and\ \bibinfo {author} {\bibfnamefont {D.~R.}\ \bibnamefont {Reichman}},\ }\bibfield  {title} {\enquote {\bibinfo {title} {Dynamic length scales in glass-forming liquids: An inhomogeneous molecular dynamics simulation approach},}\ }\href {\doibase 10.1021/jp4035419} {\bibfield  {journal} {\bibinfo  {journal} {The Journal of Physical Chemistry B}\ }\textbf {\bibinfo {volume} {117}},\ \bibinfo {pages} {13259--13267} (\bibinfo {year} {2013})},\ \bibinfo {note} {pMID: 23883366},\ \Eprint {http://arxiv.org/abs/https://doi.org/10.1021/jp4035419} {https://doi.org/10.1021/jp4035419} \BibitemShut {NoStop}%
\bibitem [{\citenamefont {Götze}(2008)}]{Gotze2008}%
  \BibitemOpen
  \bibfield  {author} {\bibinfo {author} {\bibfnamefont {W.}~\bibnamefont {Götze}},\ }\href {\doibase 10.1093/acprof:oso/9780199235346.001.0001} {\emph {\bibinfo {title} {{Complex Dynamics of Glass-Forming Liquids: A Mode-Coupling Theory}}}}\ (\bibinfo  {publisher} {Oxford University Press},\ \bibinfo {year} {2008})\BibitemShut {NoStop}%
\bibitem [{\citenamefont {Janssen}(2018)}]{MCT_Janssen}%
  \BibitemOpen
  \bibfield  {author} {\bibinfo {author} {\bibfnamefont {L.~M.~C.}\ \bibnamefont {Janssen}},\ }\bibfield  {title} {\enquote {\bibinfo {title} {Mode-coupling theory of the glass transition: A primer},}\ }\href {\doibase 10.3389/fphy.2018.00097} {\bibfield  {journal} {\bibinfo  {journal} {Frontiers in Physics}\ }\textbf {\bibinfo {volume} {6}} (\bibinfo {year} {2018}),\ 10.3389/fphy.2018.00097}\BibitemShut {NoStop}%
\bibitem [{\citenamefont {Biroli}\ \emph {et~al.}(2006)\citenamefont {Biroli}, \citenamefont {Bouchaud}, \citenamefont {Miyazaki},\ and\ \citenamefont {Reichman}}]{Biroli2006}%
  \BibitemOpen
  \bibfield  {author} {\bibinfo {author} {\bibfnamefont {G.}~\bibnamefont {Biroli}}, \bibinfo {author} {\bibfnamefont {J.-P.}\ \bibnamefont {Bouchaud}}, \bibinfo {author} {\bibfnamefont {K.}~\bibnamefont {Miyazaki}}, \ and\ \bibinfo {author} {\bibfnamefont {D.~R.}\ \bibnamefont {Reichman}},\ }\bibfield  {title} {\enquote {\bibinfo {title} {Inhomogeneous mode-coupling theory and growing dynamic length in supercooled liquids},}\ }\href {\doibase 10.1103/PhysRevLett.97.195701} {\bibfield  {journal} {\bibinfo  {journal} {Phys. Rev. Lett.}\ }\textbf {\bibinfo {volume} {97}},\ \bibinfo {pages} {195701} (\bibinfo {year} {2006})}\BibitemShut {NoStop}%
\end{thebibliography}%

\end{document}